\documentclass[fleqn,usenatbib]{mnras}

\usepackage{newtxtext,newtxmath}

\usepackage[T1]{fontenc}

\DeclareRobustCommand{\VAN}[3]{#2}
\let\VANthebibliography\thebibliography
\def\thebibliography{\DeclareRobustCommand{\VAN}[3]{##3}\VANthebibliography}

\usepackage{graphicx}	
\usepackage{amsmath}	
\usepackage{tabularx}
\usepackage[flushleft]{threeparttable}
\usepackage{xfrac}

\newcommand{\Msol}{M_{\odot}}
\newcommand{\ang}{\textup{\AA}}

\title[Actinide signatures in low $Y_e$ kilonova ejecta]{Actinide signatures in low electron fraction kilonova ejecta}

\author[Quentin Pognan et al.]{
Quentin Pognan$^{1,2}$\thanks{E-mail: quentin.pognan@aei.mpg.de}, 
Meng-Ru Wu$^{3,4,5}$,
Gabriel Mart{\'i}nez-Pinedo$^{6,7}$,
Ricardo Ferreira da Silva$^{8,9}$, \newauthor
Anders Jerkstrand$^{2}$,
Jon Grumer$^{10}$,
Andreas Fl{\"o}rs$^{6}$
\\
$^{1}$Max Planck Institute for Gravitational Physics (Albert Einstein Institute), Am M\"{u}hlenberg 1, Potsdam-Golm, 14476, Germany\\
$^{2}$Oskar Klein Center, Department of Astronomy, Stockholm University Albanova, SE-10691, Stockholm, Sweden\\
$^{3}$Institute of Physics, Academia Sinica Taipei, 115201, Taiwan \\
$^{4}$Institute of Astronomy and Astrophysics, Acadamia Sinica Taipei, 106319, Taiwan \\
$^{5}$Physics Division, National Center for Theoretical Sciences, Taipei 106319, Taiwan \\
$^{6}$GSI Helmholtzzentrum f{\"u}r Schwerionenforschung Planckstra{\ss}e 1, D-64291, Darmstadt, Germany \\
$^{7}$Institut f{\"u}r Kernphysik (Theoriezentrum), Technische Universit{\"a}t Darmstadt Schlossgartenstra{\ss}e 2, D-64289, Darmstadt, Germany \\
$^{8}$Laboratório de Instrumentação e Física Experimental de Partículas (LIP) \\ Av. Prof. Gama Pinto 2, 1649-003 Lisboa, Portugal \\
$^{9}$Faculdade de Ciências da Universidade de Lisboa \\ Rua Ernesto de Vasconcelos, Edifício C8, 1749-016, Lisboa, Portugal \\
$^{10}$Theoretical Astrophysics, Department of Physics and Astronomy Uppsala University, Box 516, 75120 Uppsala, Sweden
}

\date{Accepted XXX. Received YYY; in original form ZZZ}

\pubyear{\the\year{}}

\begin{document}
\label{firstpage}
\pagerange{\pageref{firstpage}--\pageref{lastpage}}
\maketitle

\begin{abstract}
Neutron star (NS) mergers are known to produce heavy elements through rapid neutron capture (r-process) nucleosynthesis. Actinides are expected to be created solely by the r-process in the most neutron rich environments. Confirming if NS mergers provide the requisite conditions for actinide creation is therefore central to determining their origin in the Universe. Actinide signatures in kilonova (KN) spectra may yield an answer, provided adequate models are available in order to interpret observational data. In this study, we investigate actinide signatures in neutron rich merger ejecta. We use three ejecta models with different compositions and radioactive power, generated by nucleosynthesis calculations using the same initial electron fraction ($Y_e = 0.15$) but with different nuclear physics inputs and thermodynamic expansion history. These are evolved from 10 - 100 days after merger using the \textsc{sumo} non-local thermodynamic equilibrium (NLTE) radiative transfer code. We highlight how uncertainties in nuclear properties, as well as choices in thermodynamic trajectory, may yield entirely different outputs for equal values of $Y_e$. We consider an actinide-free model and two actinide-rich models, and find that the emergent spectra and lightcurve evolution are significantly different depending on the amount of actinides present, and the overall decay properties of the models. We also present potential key actinide spectral signatures, of which doubly ionized $_{89}$Ac and $_{90}$Th may be particularly interesting as spectral indicators of actinide presence in KN ejecta.
\end{abstract}




\section{Introduction} 
\label{sec:intro}

Compact object mergers involving at least one neutron star (NS), either with a second NS in a binary system (BNS) or with a black hole (BHNS), are broadly accepted to be sites of rapid neutron capture (r-process) nucleosynthesis \citep{Lattimer.Schramm:74,Lattimer.Schramm:76,Symbalisty.Schramm:82,Eichler.etal:89,Rosswog.etal:99,Freiburghaus.etal:99}, with observations of the kilonova (KN) AT2017gfo confirming that the electromagnetic (EM), radioactively powered counterparts to these events broadly follow theoretical predictions \citep[e.g.][]{Abbott.etal:17,Cowperthwaite.etal:17,Kasen.etal:17,Smartt.etal:17,Villar.etal:17}. AT2017gfo has been investigated in great detail, with both lightcurve (LC) \citep[e.g.][]{Tanaka.etal:18,Wollaeger.etal:18,Bulla.etal:19,Kawaguchi.etal:21} and spectral modelling indicating the presence of a broad range of r-process elements \citep[e.g.][]{Watson.etal:2019,Domoto.etal:21,Gillanders.etal:21,Domoto.etal:22,Gillanders.etal:22,Hotokezaka.etal:23,Sneppen.Watson:23,Tarumi.etal:23}. Since AT2017gfo, two further KN candidates have been observed following the detection of long gamma-ray burst (lGRB) afterglows: GRB 211211A \citep{Rastinejad.etal:22} and GRB 230307A \citep{Levan.etal:24}, though these lack the complete broadband observations of AT2017gfo, and only the latter has spectral data.

Through the study of NS mergers and subsequent KNe, the origin of r-process elements in the Universe may be investigated. Of particular interest is the creation of the heaviest known elements within the third r-process peak ($Z \geq 72$) and beyond to the actinides. These require extremely neutron rich conditions (electron fractions $Y_e \lesssim 0.2$) within the ejecta in order to be synthesised \citep[e.g.][]{Goriely.etal:11,Korobkin.etal:12,Wanajo.etal:14,Wu.etal:16,Nedora.etal:21,Wu.Banerjee:22,Holmbeck.etal:23}. This criterion is broadly expected to be met in part of the dynamic ejecta of NS mergers, depending on the impact of weak interactions \citep[e.g.][]{Wanajo.etal:14,Goriely.etal:15,Fujibayashi.etal:20b,Nedora.etal:21,Kullmann.etal:22,Fujibayashi.etal:23}. Based on the nature of the central remnant, e.g. a promptly collapsed BH or short-lived (10s of ms) NS remnant, the requisite ejecta conditions may also be produced by accretion disc outflows \citep[see e.g.][]{Shibata.Hotokezaka:19,Wanajo.etal:22,Kawaguchi.etal:24}. In order to fully determine the origin of actinides, and to answer the question of whether NS mergers are the sole astrophysical sites providing sufficiently neutron-rich conditions to create them, it is necessary to understand their observational signatures in KNe.

While observational data are relatively lacking, theoretical models predict a significant diversity in NS mergers and their associated KNe. Notably, LCs and emergent spectra are expected to vary depending on merger properties, such as component masses, spins, inclination to the observer. In addition, modelling depends on the nuclear equation of state (EoS), as well as diverse nuclear quantities related to the r-process nucleosynthesis occurring within the ejecta \citep[e.g.][]{Barnes.etal:21,Zhu.etal:21,Bulla:23,Kullmann.etal:23,Sarin.Rosswog:24}. The models produced for the analysis of KNe are therefore equally varied, with their intrinsic accuracy depending on the uncertainty in the aforementioned quantities, as well as the physics included in the modelling. 

Of critical importance for KN modelling are the elemental composition and radioactive power of the merger ejecta, which arise from the initial r-process nucleosynthesis, and subsequent radioactive decay of unstable isotopes. As these decays power the KN, the details pertaining to specific interactions of various decay products within the ejecta directly affect the emergent LC and spectra, and thus have powerful diagnostic potential if properly understood. The radioactive power and composition are often related to the neutron abundance of the ejecta by the electron fraction $Y_e$, with lower values signifying a more neutron rich environment \citep[see e.g.][for a review on r-process nucleosynthesis]{Cowan.etal:21}. 

The parametrization of power and composition by $Y_e$ can be extended to calculations of expansion opacities, which are then used to constrain groups of elements present in KN ejecta following the evolution of broadband LCs \citep[e.g.][]{Kasen.etal:17,Tanvir.etal:17,Villar.etal:17,Banerjee.etal:20,Tanaka.etal:20,Domoto.etal:21,Domoto.etal:22,Banerjee.etal:24}. The identification of individual elements and species requires detailed spectral analysis \citep[e.g.][]{Watson.etal:2019,Domoto.etal:21,Gillanders.etal:21,Domoto.etal:22,Gillanders.etal:22,Sneppen.Watson:23,Tarumi.etal:23}. In both cases, models making use of atomic data (e.g. at least energy levels and radiative transitions), as well as ejecta compositions and power, typically from a general formula or taken from nuclear network outputs, are required in order to interpret observational data. 

The case of low $Y_e$, actinide bearing ejecta is particularly challenging due to the large degree of uncertainty associated with such neutron rich conditions. From the nuclear perspective, many studies have shown that synthesised abundances of actinide species are poorly constrained due to unknown nuclear quantities such as nuclear masses, $\beta$-decay rates, decay channel branching ratios, and nuclear fission rates \citep[e.g.][]{Mendoza.etal:15,Wu.etal:19,Giuliani.etal:21,Lemaitre.etal:21,Zhu.etal:21,Wu.Banerjee:22}. This leads to large variations in predicted abundances, as well as radioactive power, both of which are key ingredients in the modelling of KNe. Furthermore, the atomic structure of actinide species is typically poorly known, due to a lack of experimental atomic data, and the complexity of actinide species with partially filled, open f-shell configurations \citep[e.g.][]{Even.etal:20,Tanaka.etal:20,Flors.etal:23,Fontes.etal:23,Domoto.etal:24}.

In this work, we build on previous studies highlighting the observational impacts of uncertainties in the aforementioned quantities. We approach this issue from both a LC and spectral modelling perspective, making use of non-local thermodynamic equilibrium (NLTE) radiative transfer simulations in order to investigate potential actinide signatures from $\sim 10 - 100$ days after merger. We employ three models that have identical ejecta profiles and electron fractions of $Y_e = 0.15$, but whose compositions and radioactive power are taken from different nuclear network calculations. The usage of NLTE radiative transfer simulations allows us to highlight potential actinide signatures in low $Y_e$ ejecta to late times in a self-consistent manner. 

In Section \ref{sec:models} we present the models used in this study, as well as the radiative transfer simulation. We continue in Section \ref{sec:thermo_results} by presenting the temperature and ionization structure evolution of these models, highlighting the differences which lead to to spectral and LC variations as discussed in Section \ref{sec:spectral_results}. Finally, we summarise our findings and present future directions in Section \ref{sec:discusssion}.

\section{Ejecta models and radiative transfer simulation}
\label{sec:models}

\subsection{Ejecta models and compositions}

We adopt three different models of nuclear composition and associated radioactive power from \citet{Barnes.etal:16} and \citet{Wanajo.etal:14}. The calculations were computed using different thermodynamic trajectories and nuclear inputs, such as mass models and decay rates, but with the same electron fraction value of $Y_e=0.15$. The first two models, named `B16A' and `B16B',  correspond to those computed using the same nuclear reaction network, employing different sets of nuclear ground-state masses from the Duflo-Zuker \citep[DZ31][]{Duflo.Zuker:95} and the Finite Range Droplet Model \citep[FRDM][]{FRDM} respectively. The underlying thermodynamic trajectory and all other nuclear reaction rates are the same for these two models. The third model named `W14' used a different nuclear reaction network supplied with reaction rates that are all different from those used in the B16A/B inputs \citep[see][for details pertaining to these calculations]{Wanajo.etal:14,Barnes.etal:16}.

\begin{figure*}
    \centering
    \includegraphics[width=0.8\linewidth]{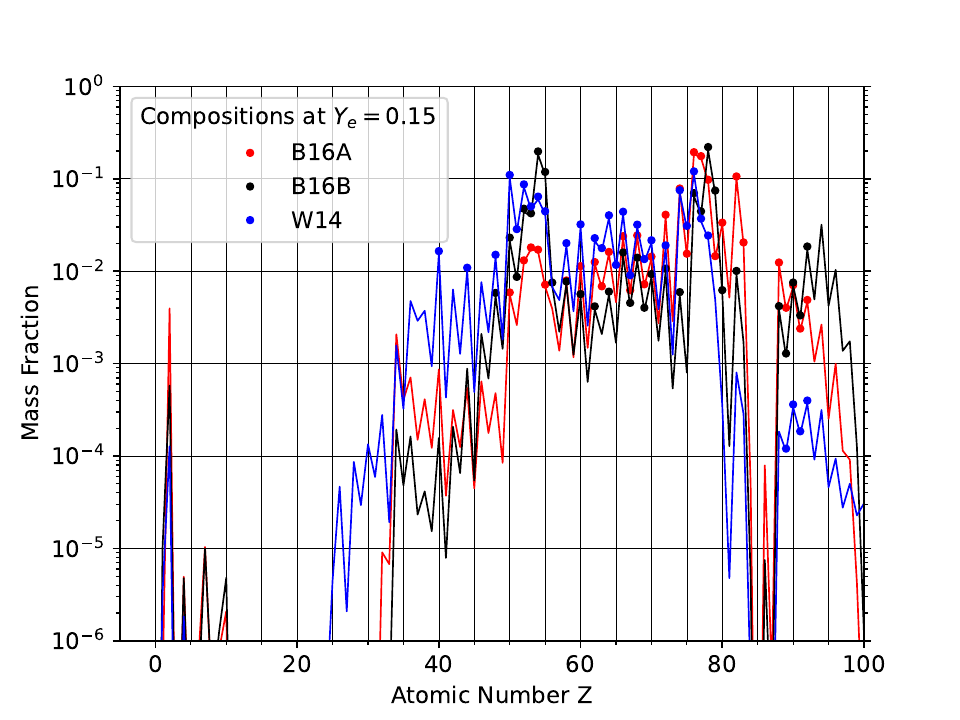}
    \caption{Compositions of the three models at 25 days (B16B, W14) and 27 days (B16A). Elements that are actively modelled in the radiative transfer simulation are marked by points (see text for explanation).}
    \label{fig:comp_compare}
\end{figure*}

\begin{table}
    \caption{Model compositions by mass fractions, with total lanthanide and actinide mass fractions. B16B and W14 are at 25 days, while B16A is at 27 days. 
    \label{tab:compositions}}
    \centering
    \setlength\tabcolsep{0.4cm}
    \begin{tabular}{cccc}
    \hline \hline
    Element & B16A & B16B & W14 \\
    \hline 
    $_{40}$Zr & - & - & 0.0165 \\
    $_{44}$Ru & - & - & 0.0109 \\
    $_{48}$Cd & - & 0.0058 & 0.0151 \\
    $_{50}$Sn & 0.0059 & 0.0231 & 0.1101\\
    $_{51}$Sb & - & 0.0087 & 0.0285 \\
    $_{52}$Te & 0.0131 & 0.0475 & 0.0871 \\
    $_{53}$I & 0.0181 & 0.0423 &  0.0502 \\
    $_{54}$Xe & 0.0171 & 0.1978 & 0.0642 \\
    $_{55}$Cs & 0.0072 & 0.1186 & 0.0445 \\
    $_{56}$Ba & - & 0.0075 & - \\
    $_{58}$Ce & 0.0079 & 0.0077 & 0.0201 \\
    $_{60}$Nd & 0.0113 & 0.0057 & 0.0321 \\
    $_{62}$Sm & 0.0126 & 0.0042 & 0.0228 \\
    $_{63}$Eu & 0.0069 & - & 0.0178 \\
    $_{64}$Gd & 0.0162 & 0.0060 & 0.0402 \\
    $_{65}$Tb & - & - & 0.0117 \\
    $_{66}$Dy & 0.0239 & 0.0159 & 0.0440 \\
    $_{67}$Ho & 0.0062 & 0.0045 & 0.0091 \\
    $_{68}$Er & 0.0244 & 0.0140 & 0.0319 \\
    $_{69}$Tm & 0.0072 & 0.0040 & 0.0135 \\
    $_{70}$Yb & 0.0144 & 0.0093 & 0.0216 \\
    $_{72}$Hf & 0.0408 & 0.0107 & 0.0190 \\
    $_{74}$W &  0.0785 & 0.0059 & 0.0753 \\
    $_{75}$Re & 0.0155 & - & 0.0310 \\
    $_{76}$Os & 0.1941 & 0.0700 & 0.1204 \\
    $_{77}$Ir & 0.1759 & 0.0445 & 0.0371 \\
    $_{78}$Pt & 0.0975 & 0.2205 & 0.0243 \\
    $_{79}$Au & 0.0146 & 0.0746 & - \\
    $_{80}$Hg & 0.0335 & 0.0062 & - \\
    $_{82}$Pb & 0.1060 & 0.0101 & - \\
    $_{83}$Bi & 0.0205 & - & - \\
    $_{88}$Ra & 0.0124 & 0.0042 & - \\
    $_{89}$Ac & 0.0040 & 0.0013 & 0.0001 \\
    $_{90}$Th & 0.0070 & 0.0075 & 0.0004 \\
    $_{91}$Pa & 0.0024 & 0.0033 & 0.0002 \\
    $_{92}$U & 0.0049 & 0.0185 & 0.0004 \\ \hline
    $X_{\rm{La}}$ & 0.1310 & 0.0713 & 0.2648 \\
    $X_{\rm{Ac}}$ & 0.0183 & 0.0306 & 0.0011 \\
    \hline \hline
    \end{tabular}
\end{table}

\begin{figure}
    \centering
    \includegraphics[trim={0.4cm 0cm 0.4cm 0.2cm},width=1.0\linewidth]{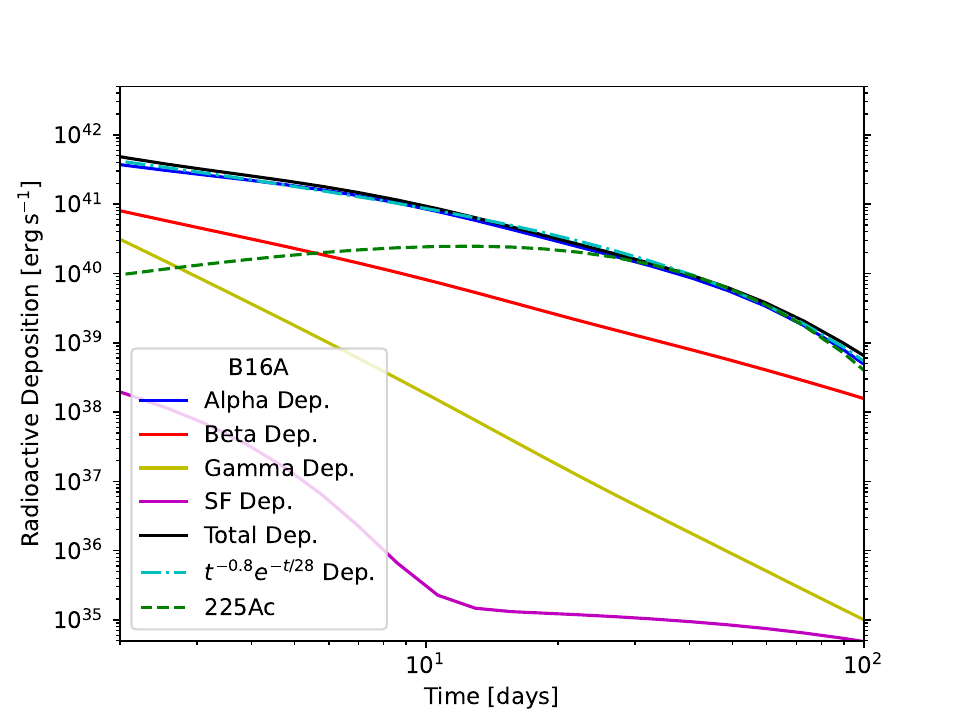} \\
    \includegraphics[trim={0.4cm 0cm 0.4cm 0.2cm},width=1.0\linewidth]{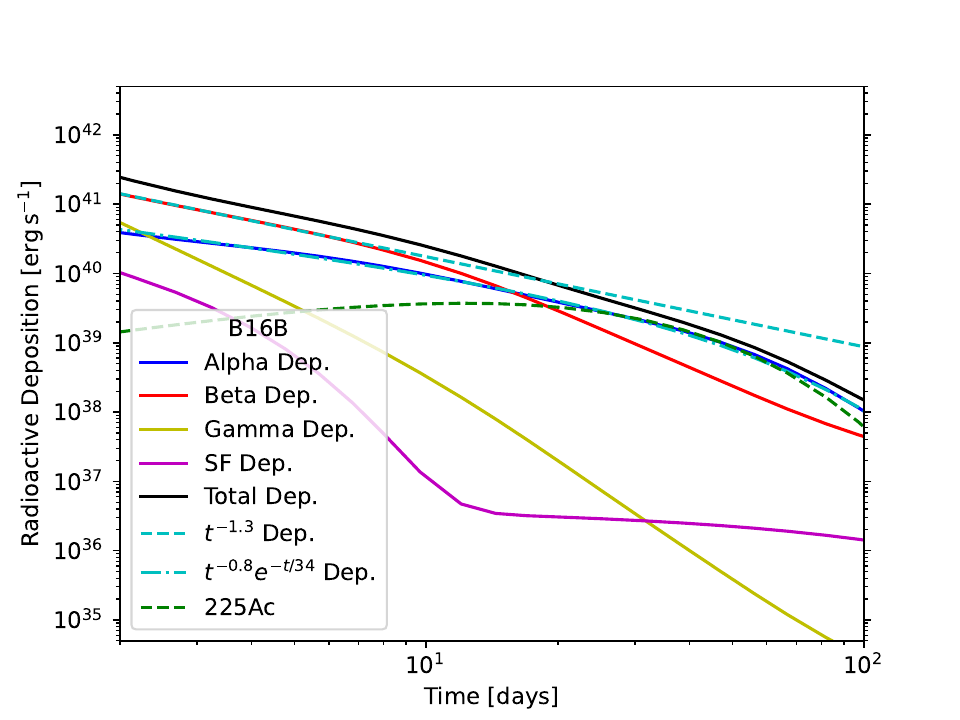} \\
    \includegraphics[trim={0.4cm 0cm 0.4cm 0.2cm},width=1.0\linewidth]{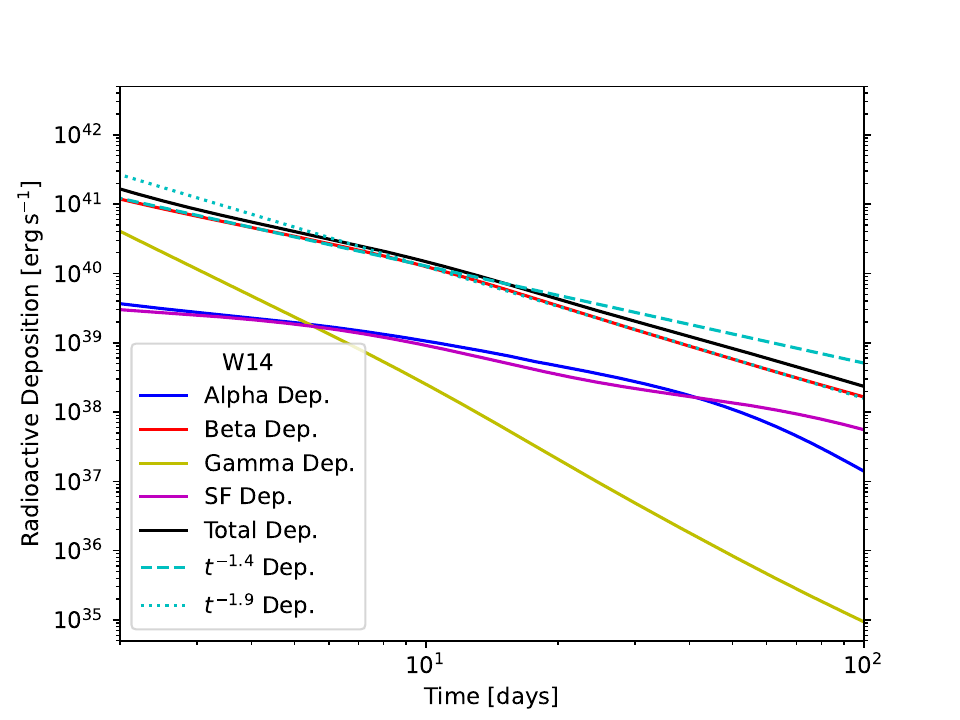}
    \caption{Energy depositions, including thermalization, by decay product for each model. Fits to the dominant sources of energy deposition in the models are shown in cyan. The green dashed line in the top and middle panels shows the deposition arising from the decay chain of $^{225}$Ac. Note that the simulation range is from 10 - 100 days.}
    \label{fig:dep_compare}
\end{figure}

The ejecta models used in this study are all single-dimensional, composed of 5 radial velocity zones spanning $v_{\rm{ej}} = 0.05c - 0.30c$, increasing in steps of $0.05c$. The total ejecta mass is $M_{\rm{ej}} = 0.05 \Msol$, and follows a density profile of $\rho \propto v^{-4}$. The models are all placed at 40 Mpc from the observer, similar to the distance inferred for AT2017gfo \citep[e.g.][]{Abbott.etal:17}. The compositions of each model are tabulated in Table~\ref{tab:compositions}. These compositions can be viewed graphically in Fig.~\ref{fig:comp_compare}, and the corresponding energy depositions in Fig.~\ref{fig:dep_compare}.

The compositions are limited to 30 elements, each with ion stages from neutral to triply ionized, due to memory constraints. The actinide species for which atomic data are available, $_{89}$Ac - $_{92}$U, are always selected, and the next 26 most abundant elements are then chosen. These abundances are taken at a fixed epoch for each model: 25 days for B16B and W14, and 27 days for B16A. This is done in order to avoid changing the model composition for each epoch, which would increase runtime and computational costs. The aforementioned epochs are chosen such that the composition reflects a rough average of the overall composition over the timespan of interest, $\sim 10 - 100$ days. From the nucleosynthesis calculations, we find that the maximal mass-fraction change in this time range occurs for $_{89}$Ac in the B16A model, which has a fraction of $\sim 0.016$ at 11 days, $\sim 0.009$ at 109 days, and $\sim 0.012$ at our chosen epoch of 27 days. Note that these values are lower than those in Table~\ref{tab:compositions}, as they are relative to \textit{all} elements from $\rm{Z} = 1 - 105$, whereas the former are relative to the 30 chosen elements. While this does introduce a small level of inconsistency between the radioactive power and composition, the inaccuracy induced by this choice is believed to be far inferior to the intrinsic uncertainty from nuclear network inputs at this low $Y_e$, as well as the thermalization formulae used to calculate the final energy deposition (see below). 

It should also be noted that the atomic mass of the elements is taken to be that of the most stable isotope, as the radiative transfer simulations do not distinguish between different isotopes of a given element. This will introduce some inconsistency in the number fractions derived from the mass fractions, as there will be variations in which isotopes are present at a given time, for a given species, due to the decay of unstable isotopes. However, as the models presented here are not sensitive enough to be greatly affected by the relatively small variations in number fractions introduced by this inconsistency, we expect the uncertainty arising from taking a constant atomic mass for each element to have a negligible impact on the emergent spectra.

From Figs.~\ref{fig:comp_compare} and~\ref{fig:dep_compare}, the different outputs of the three models for an identical electron fraction are apparent. Considering first the B16B and B16A models which only differ by ground-state nuclear masses, we find significant variations in actinide production. Notably, the B16B model predicts peak actinide abundance for $_{94}$Pu, as well as significant abundances of $_{96}$Cm and $_{98}$Cf, whereas the B16A values are markedly lower (red and black curves in Fig. \ref{fig:comp_compare}). Unfortunately, atomic data for such heavy species are lacking, such that their spectral signatures cannot be modelled directly in the radiative transfer simulation. However, these species are important starting points for energetic spontaneous fission (SF) and $\alpha$-decay channels that may dominate power production should they be present in sufficient amounts \citep[e.g.][]{Barnes.etal:16,Rosswog.etal:17,Zhu.etal:18,Wanajo:18,Wu.etal:19,Giuliani.etal:21,Holmbeck.etal:23}. The total actinide mass fraction is also significantly higher in the B16B model at $X_{\rm{Ac}} = 0.0306$ compared to the B16A model at $X_{\rm{Ac}} = 0.0183$.

Looking at Fig. \ref{fig:dep_compare}, we see that the B16A model has an energy deposition that is dominated by $\alpha$-decay across the entire 1 -- 100~day timespan. There are several key differences between $\alpha$ and $\beta$-decay that must be taken into account in order to understand the impact on energy deposition. For similar lifetimes, $\alpha$-decays are typically stronger energy sources than $\beta$-decays, as they usually involve chains of several $\alpha$-decays (e.g. 4 for $^{225}$Ac) that produce $\alpha$-particles with kinetic energies significantly higher than $\beta$-decay electrons, $\sim 6$ MeV cf. $\sim 0.5$ MeV. From this, the raw radioactive power of $\alpha$-decay is greater than that of $\beta$-decay. The energy deposition is then further augmented by the efficient thermalization of $\alpha$-particles arising from their greater energy and higher charges \citep[e.g.][]{Barnes.etal:16,Kasen.Barnes:19,Wu.etal:19}.

The $\alpha$-decay energy deposition in the two B16A/B models initially follows a power law, succeeded by an exponential decay. This is consistent with many nuclei undergoing $\alpha$-decay at early times and therefore yielding a power law relation, much like the canonical $t^{-1.4}$ power law for ensemble $\beta$-decay. At late times, we recover the specific decay chain of $\beta$-decay of $^{225}$Ra followed by $\alpha$-decay of $^{225}$Ac, as found previously in \citet{Wu.etal:19}. The B16B model has some additional late time contribution from the $\alpha$-decay of $^{253}$Es. The fact that the deposition follows the radioactive power so closely highlights how well $\alpha$-particles continue to thermalize in the bulk of the ejecta across the timespan considered here. It is also notable that the B16A model is much more energetic than the B16B model over the entirety of the 1 - 100 day range, which is expected to lead to brighter emergent luminosities.

In comparison to the B16A/B models, the W14 model produces a markedly different abundance pattern, particularly for the actinides. A very low actinide mass-fraction of $X_{\rm{Ac}}\sim 10^{-3}$ is yielded, approximately an order of magnitude less than the B16A and B16B calculations. Conversely, a higher lanthanide mass fraction is obtained ($X_{\rm{La}}\sim 0.26$ cf. $\sim 0.13$ for B16A and $\sim 0.07$ for B16B), as well as significantly higher abundances of elements ranging from $_{36}$Kr to $_{50}$Sn, thus covering parts of the first and second r-process peaks. 

The difference in heavy element production and their subsequent decay is also reflected in the energy deposition of this model, which is entirely dominated by $\beta$-decay, as shown in the bottom panel of Fig. \ref{fig:dep_compare}. The early deposition follows the canonical $\dot{q}_{\rm{tot}} \sim t^{-1.4}$ power law, and drops to $\sim t^{-1.9}$ at later times. This contrasts with the predicted $t^{-2.8}$ post-thermalization break scaling \citep[][]{Kasen.Barnes:19,Waxman.etal:19,Hotokezaka.etal:20}. This is likely due to the layered nature of the ejecta, such that the inner layers are still thermalizing efficiently at late times, while the outer layers are not. This leads to an overall deposition that follows a power law steeper than the radioactive power generation at $t^{-1.4}$, but shallower than the complete post-thermalization break deposition at $t^{-2.8}$.

The W14 model also has a small spontaneous fission contribution at $\gtrsim 70$ days, which yet remains subdominant with respect to $\beta$-decay. A consequence of this, is that the W14 model has the lowest overall energy deposition for most of the epochs considered here, only becoming slightly more energetic than the B16B model at late times when the fission contribution comes in. The B16A model remains by far the most energetic over the whole period by a factor of $\sim 5$ more so than the other two models. 

The above discussion highlights the importance of using time-dependent, radioactive energy deposition rates with consideration of the distribution of the various decay products as determined from network outputs, in order to fully capture the dependence on nuclear properties \citep[e.g.][]{Barnes.etal:21}. These effects are difficult to capture by analytical heating formulae despite the inclusion of multiple terms \citep[e.g.][]{Rosswog.Korobkin:24}. In addition, the thermalization of decay products plays a fundamental role in the energy deposition, a subject which has been greatly studied \citep[][]{Barnes.etal:16,Wollaeger.etal:18,Kasen.Barnes:19,Waxman.etal:19,Hotokezaka.etal:20,Zhu.etal:21}. In this work, the analytical thermalization equations from \citet{Kasen.Barnes:19} are used for $\alpha$-decay particles, $\beta$-decay electrons/positrons, and $\gamma$-rays, with an additional input to the $\alpha$ and $\beta$-decay treatments from \citet{Waxman.etal:19}, while the prescription from \citet{Barnes.etal:16} is employed for fission products, regardless of initial injection energy \citep[see Appendix A of][for a detailed description of the employed formulae]{Pognan.etal:23}. While this treatment may not be the most advanced, the focus of this work is not the uncertainty present within thermalization physics, but rather the impact of uncertainties in the nuclear network inputs, i.e. the raw radioactive power, on potential actinide signatures in KNe. As such, an equal application of a relatively simple thermalization prescription is deemed to be satisfactory for the goals of this study.

\subsection{Radiative transfer simulation}
\label{subsec:RT}

The radiative transfer simulations employ the spectral synthesis code \textsc{sumo} \citep{Jerkstrand.etal:11,Jerkstrand.etal:12}. \textsc{sumo} is a NLTE Monte Carlo code originally designed for nebular phase supernovae and adapted to model KNe, with the most recent updates for KN modelling described in \citet{Pognan.etal:23}. The three models are evolved over a timespan of $\sim 10 - 100$ days, in time-dependent NLTE mode \citep[as initially described in][]{Pognan.etal:22a}. 

Compared to previous work, thermal collision strengths have been updated based on the work of \citet{Bromley.etal:23}, which found that the van Regemorter \citep[][for allowed transitions]{Regemorter:62} and Axelrod \citep[][for forbidden transitions]{Axelrod:80} prescriptions systematically underestimate collision strengths for neutral to doubly ionized $_{78}$Pt when compared to values found by detailed R-matrix calculations. Though this study was for a single r-process element, we take it as indicative of collision strength underestimation for all our r-process species, as other works for specific transitions have also found similar results \citep[e.g. the Te \textsc{iii} 2.1$\mu$m line][]{Madonna.etal:18,Hotokezaka.etal:23}. In order to broadly correct for this effect, we apply a simple scaling factor to our calculations using the van Regemorter and Axelrod prescriptions, the exact value of which depends on ionization stage and temperature solution, but lying in the range of $1.5 - 10$ (see Appendix \ref{app:thermal_coll} for precise values). The scaling values were found by taking the average offset between the values calculated for Pt \textsc{i} - \textsc{iii} by the Van Regemorter and Axelrod formulae, and R-matrix calculations \citep[see Figure 3 of][for a visualisation]{Bromley.etal:23}. While this may lead to overestimation of collision strengths for certain transitions, it is impossible to quantify this effect, or even identify for which transitions this will occur until more atomic data for this process are available. Generally, increased collision strengths may lead to enhanced cooling as thermal collisions become more efficient. 

Radiative transfer simulations of KNe must in general make use of atomic data, particularly atomic energy levels and transitions (wavelengths and A-values) in order to model the r-process species interacting in the ejecta. It is well known that the sparsity of atomic data for the majority of r-process species poses a particular challenge to KN modelling, which is found to be greater when addressing complex, line-rich species such as lanthanides and actinides \citep[e.g.][]{Even.etal:20,Tanaka.etal:20,Flors.etal:23,Fontes.etal:23}. Due to the lack of experimental atomic data, theoretical data sets calculated by atomic physics codes are often used.

The atomic data used in this study comes from calculations done using the Flexible Atomic Code \citep[\textsc{fac}][version from April 21st, 2021\footnote{Corresponding to commit id \textsc{ac5f58c}.}]{Gu:08}, and are identical to those used in \citet{Pognan.etal:23}, the purpose of which was completeness rather than detailed accuracy for key species. The configurations used for the \textsc{fac} calculations can be found in the appendix of \citet{Pognan.etal:23} \footnote{See erratum of \citet{Pognan.etal:23} concerning the configurations of U\,\textsc{iv}.}, though it should be noted that the development of the methods implemented in \textsc{fac} has caused variations in the resultant atomic data over the years from seemingly similar models. This should be considered when making detailed comparisons between data sets computed with different versions.

The NLTE radiative transfer simulations also make use of diverse cross-sections and rates (e.g. photoionization, recombination etc.) for the various processes included in the modelling. For r-process elements, data relating to these quantities are similarly lacking, such that fitting formulae or estimations are used instead. The details of the values used to model these diverse processes are described in \citet{Pognan.etal:23}, with only the aforementioned treatment of thermal collision strengths having been modified for this study. 

\section{Thermodynamic Evolution}
\label{sec:thermo_results}

\begin{figure*}
    \centering
    \includegraphics[width=0.49\linewidth]{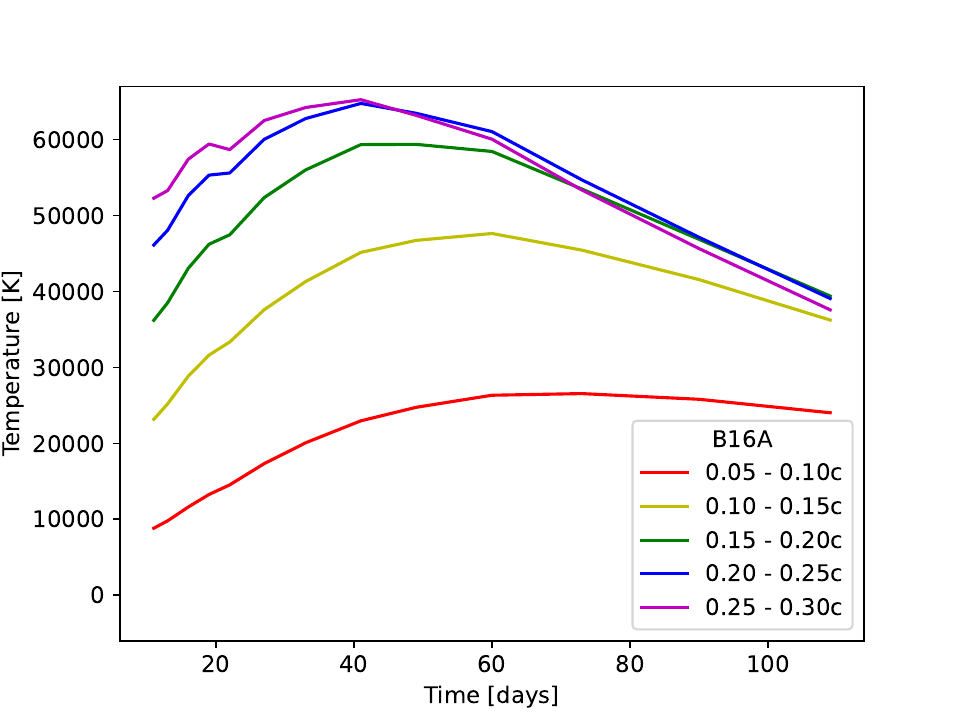}
    \includegraphics[width=0.49\linewidth]{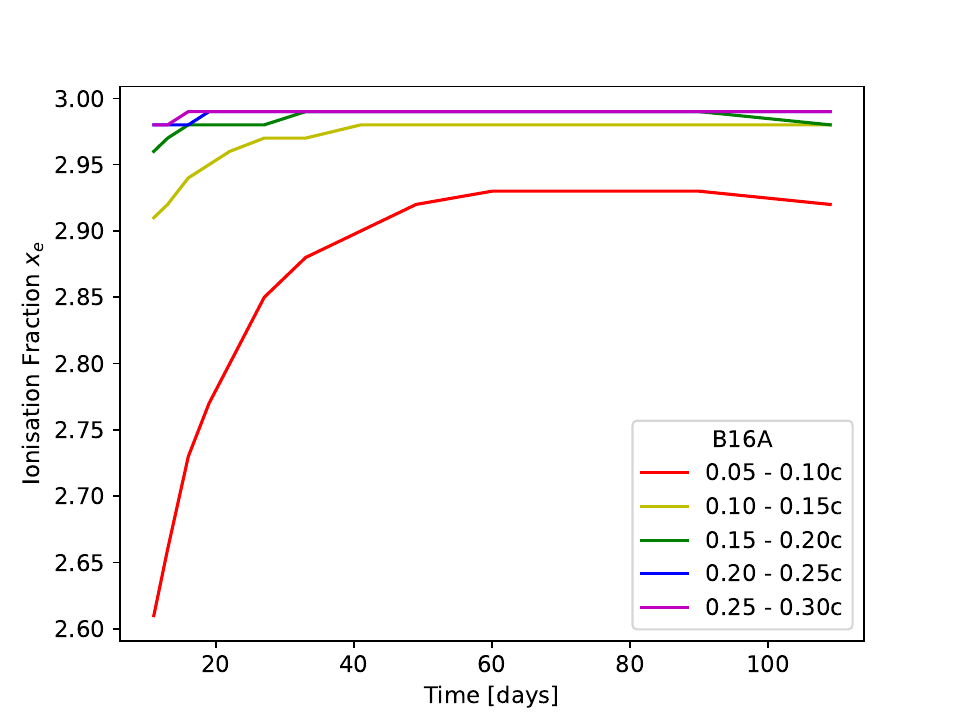} 
    \includegraphics[width=0.49\linewidth]{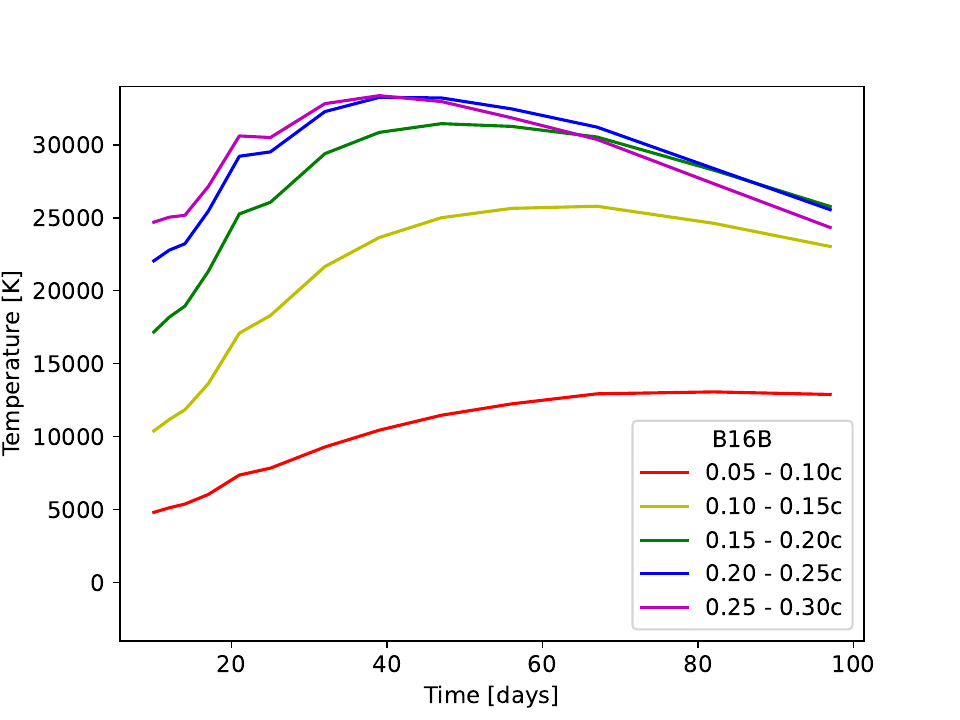}
    \includegraphics[width=0.49\linewidth]{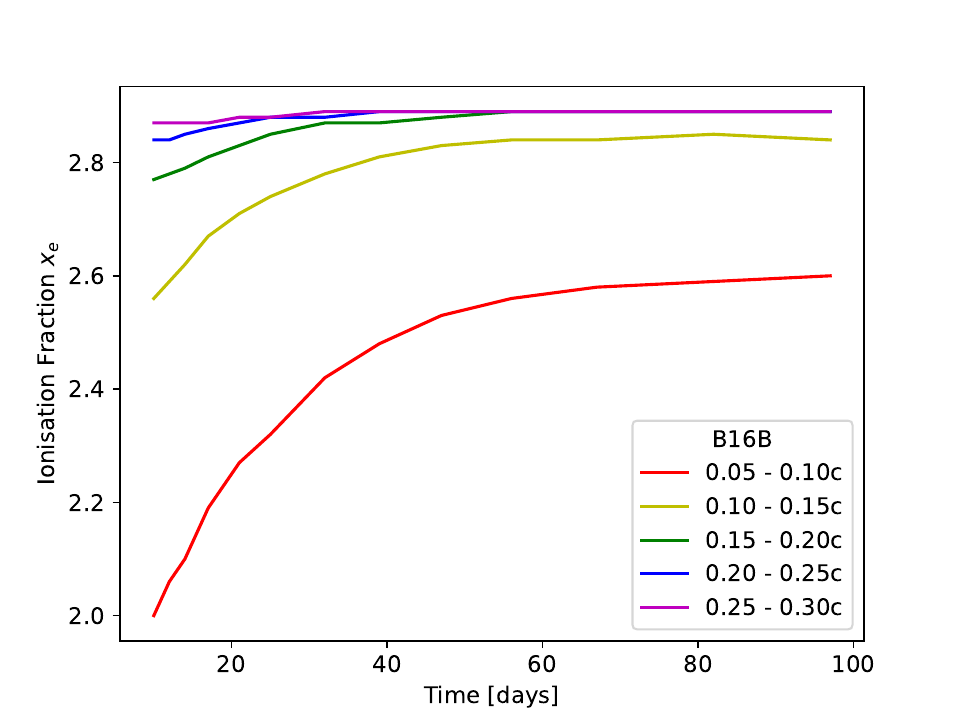} 
    \includegraphics[width=0.49\linewidth]{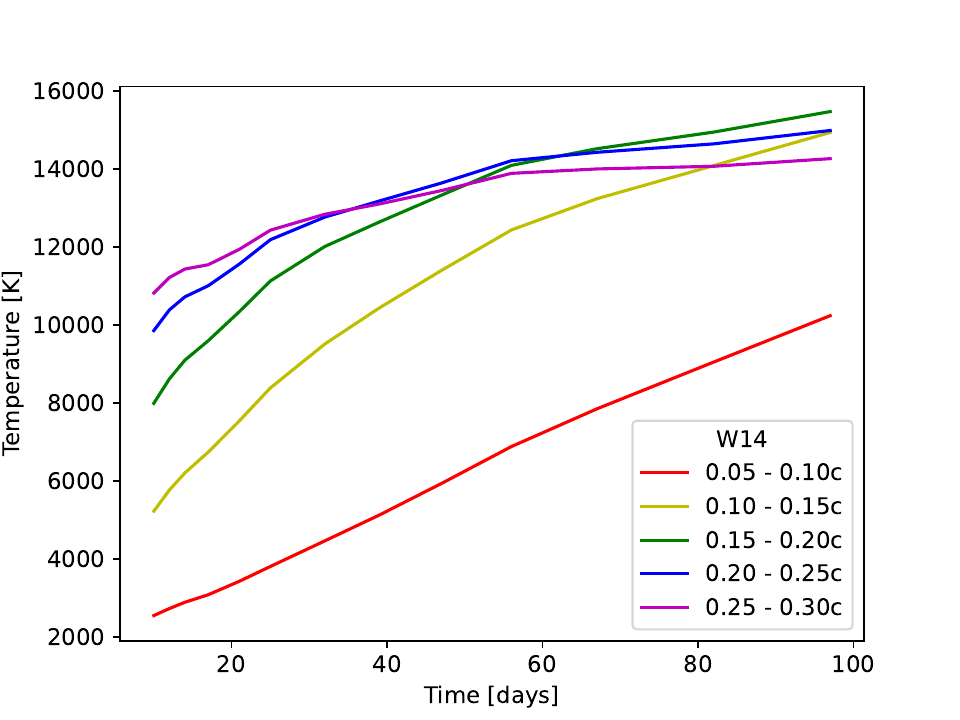}
    \includegraphics[width=0.49\linewidth]{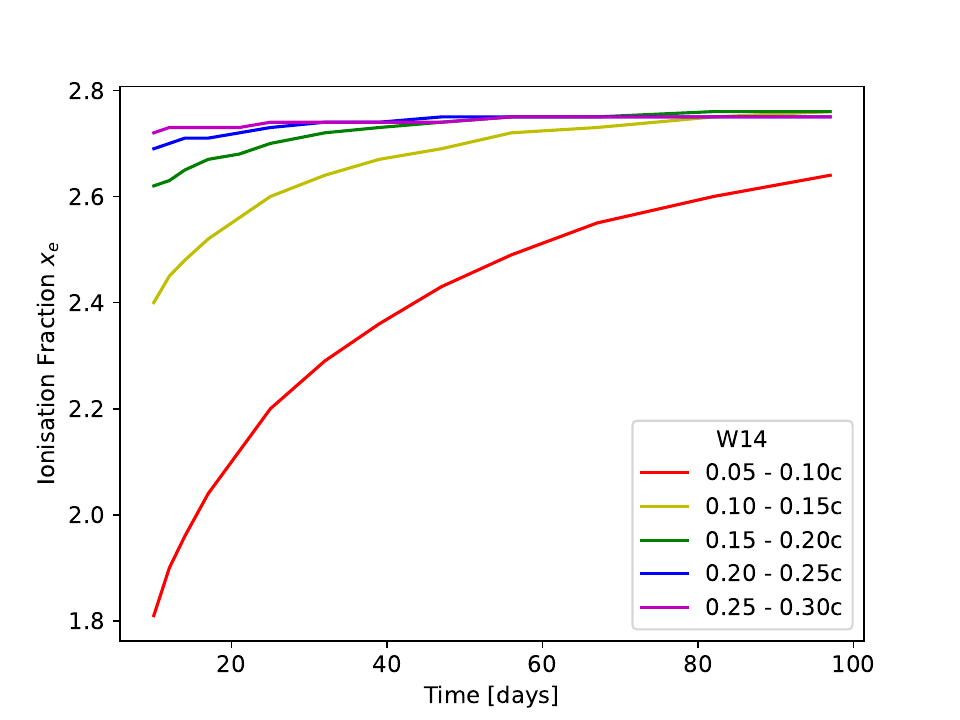}
    \caption{Temperature (left-hand panels) and ionization structure (right-hand panels) evolution of the three models, for each model zone. The ionization structure is characterized by electron fraction $x_e$, representing the number of free electrons per ion. Note that this is different from the compositional electron fraction $Y_e$.}
    \label{fig:thermo}
\end{figure*}

We consider first the temperature and ionization structure evolution of the three models, which can be viewed in Fig. \ref{fig:thermo}. Considering first the ionization structures shown in the right-hand panels, we see that all models exhibit remarkably similar evolution and structure, where the degree of ionization follows the overall energy deposition of the model, i.e. B16A is the most ionized, and W14 the least. A stratification is broadly found in every model, where the inner layers are typically less ionized than the outer layers, though some minor inversions do occur (e.g. in the W14 model at very late times). The low density in these outer layers inhibits both recombination and ionization processes, with recombination being more affected \citep[see][for an analysis of ionization and recombination rates]{Pognan.etal:23}, thus usually leading to greater degrees of ionization. 

In every model, the outermost three layers have an exceptionally flat ionization evolution, which is the result of time-dependent effects in the low density regime \citep{Pognan.etal:22a,Pognan.etal:23}. In this regime, the efficiency of ionization and recombination is so low that a structure freeze-out is rapidly experienced, as the time-scales associated to these processes are comparable to, or even longer than the evolutionary time. 

Considering now the temperature evolution, it is immediately noticeable that the B16A/B models are broadly similar, though the former is approximately twice as hot as a result of the greater energy deposition. Both models show an initially increasing temperature, which experiences a downturn and then begins to cool. The exact time at which this happens is density dependent, with the lower density, outer layers experiencing the downturn first around 40 days, while the innermost zone has a slower transition, around 60 days for B16A, and cooling has yet to fully begin in the innermost layer of the B16B model. It is also noteworthy that the slope of the temperature drop is greater for the outer layers, such that the outermost layer is no longer the hottest at 100 days. The slope of the outermost layer of the B16A model is also greater than that of the B16B model, which follows from enhanced line-cooling and adiabatic cooling at hotter temperatures. 

In contrast, the W14 model follows more closely the evolution predicted by \citet{Hotokezaka.etal:21,Pognan.etal:22a}, that is, a monotonically increasing temperature with time. It should be noted however, that the overall energy deposition does not follow the predicted post-thermalization break relation of $t^{-2.8}$, as explained in Section \ref{sec:models}. As such, the exact temperature evolution for each zone does not necessarily follow the predicted $T \sim t^{0.2}$ relation \citep[][]{Hotokezaka.etal:21}, and it is clearly seen that the temperature for the innermost zone is still rising significantly faster than the outer zones which are inefficiently thermalizing $\beta$-decay electrons. The outermost layers of the W14 model also exhibit a temperature flattening due to time-dependent effects. Notably, adiabatic cooling plays a significant role at these late times, contributing a maximum of $\sim 30$ per cent to the overall cooling in the outermost layer. 

The adiabatic cooling contribution is comparatively smaller for the B16A model, only reaching $\sim 11$ per cent, whereas it is much more significant in the B16B model at $\sim 47$ per cent. It should be noted however, that these values represent the contribution to total cooling; the adiabatic cooling in the B16A model is still physically greater than in the B16B model as the total cooling is greater, driven by hotter temperatures. Despite its notable contribution to the total cooling, adiabatic cooling is not the sole driving factor behind the temperature downturn. Indeed, the prediction of a monotonically increasing temperature assumes that $\beta$-decay dominates, and that the deposited energy from this decay follows a temporal evolution of $E_{\rm{dep}} \propto t^{-2.8}$ after the thermalization efficiency break. However, these two models are not $\beta$-decay dominated at these late times, but instead $\alpha$-decay dominated (see Fig. \ref{fig:dep_compare}). 

As described previously in Section \ref{sec:models}, the late-time $\alpha$-decay energy comes mostly from the $\alpha$-decay of $^{225}$Ac, and follows more closely an exponential decay law given by the lifetime of $^{225}$Ra (21.5 days) that $\beta$-decays to $^{225}$Ac, as opposed to the ensemble decay power law seen for $\beta$-decay. Considering the heating rate per volume, we have \citep[see equation (7) of][]{Pognan.etal:22a}:

\begin{equation}
    h(t) = M_{\rm{ej}}\dot{q}_{\rm{tot}}/ \left(\frac{4 \pi}{3} v_{\rm{ej}}^3 t^3 \right) \: \rm{erg \, s^{-1} \, cm^{-3}}
    \label{eq:heat_pervol}
\end{equation}

\noindent where $M_{\rm{ej}}$ is the ejecta mass (or zone mass for a given ejecta layer), $\dot{q}_{\rm{tot}}$ is the energy deposition including thermalization factor, $v_{\rm{ej}}$ the outer ejecta velocity (or outer zone velocity for a given ejecta layer), and $t$ is time. Assuming that $\alpha$-particles are thermalizing efficiently in bulk ejecta for the epochs of interest here, and therefore taking $\dot{q}_{\rm{tot}} \propto \rm{e}^{-t/\tau}$, we obtain a heating rate per volume scaling as $h(t) \propto t^{-3} \rm{e}^{-t/\tau}$. In the present case, we have $\tau \approx 21$~days, which is shorter than the evolutionary time-scales considered.

Conversely, the cooling rate per volume, which remains dominated by line cooling, is given by \citep[see equation (13) of][]{Pognan.etal:22a}:

\begin{equation}
    c_{\rm{line}}(t) = \sum_{i} \Lambda^{i}(T,n_e,n_i) n_e n_i \: \rm{erg \, s^{-1} \, cm^{-3}}
    \label{eq:cool_pervol}
\end{equation}

\noindent where $\Lambda^i$ is the line cooling function for a given ion species, dependent on temperature $T$, ion density $n_i$ and electron density $n_e$, in complex and non-linear ways originating from the atomic structure of the species. As the ion and electron densities scale as $t^{-3}$ in spherical homologous expansion ($V \propto v_{\rm{ej}}^3t^3$), and taking the line cooling function to be approximately independent of these densities in low enough density regimes, such as those studied here, we find a cooling per volume of $c_{\rm{line}}(t) \propto t^{-6}$. 

From the above equations, we find that the temporal evolution of the cooling at the late epochs studied here is shallower than that of the heating, once $\alpha$-decay following an exponential decay law dominates energy deposition. This is even more so when $\alpha$-decay thermalization begins to become inefficient, multiplying in a factor of $t^{-1.5}$ into the heating rate per volume, while cooling is further enhanced by adiabatic cooling scaling as $t^{-1}$ \citep[see equation (8) of ][]{Pognan.etal:23}. Therefore, late-time cooling in $\alpha$-decay dominated regimes following an exponential decay law is expected, even without consideration of time-dependent effects such as adiabatic cooling.

The exact time at which the temperature downturn occurs however, will depend on the half-life, or half-lives, of the dominating $\alpha$-decay species. Notably, a shorter half-life will lead to an earlier temperature downturn, and vice-versa for a longer half-life. As such, the details of the transition from heating to cooling dominated regimes will depend on the outputs of the nuclear networks, notably the relative abundance of $\alpha$-decaying species with respect to each other.

\begin{figure}
    \centering
    \includegraphics[trim={0.4cm 0cm 0.4cm 0.2cm},width=1.0\linewidth]{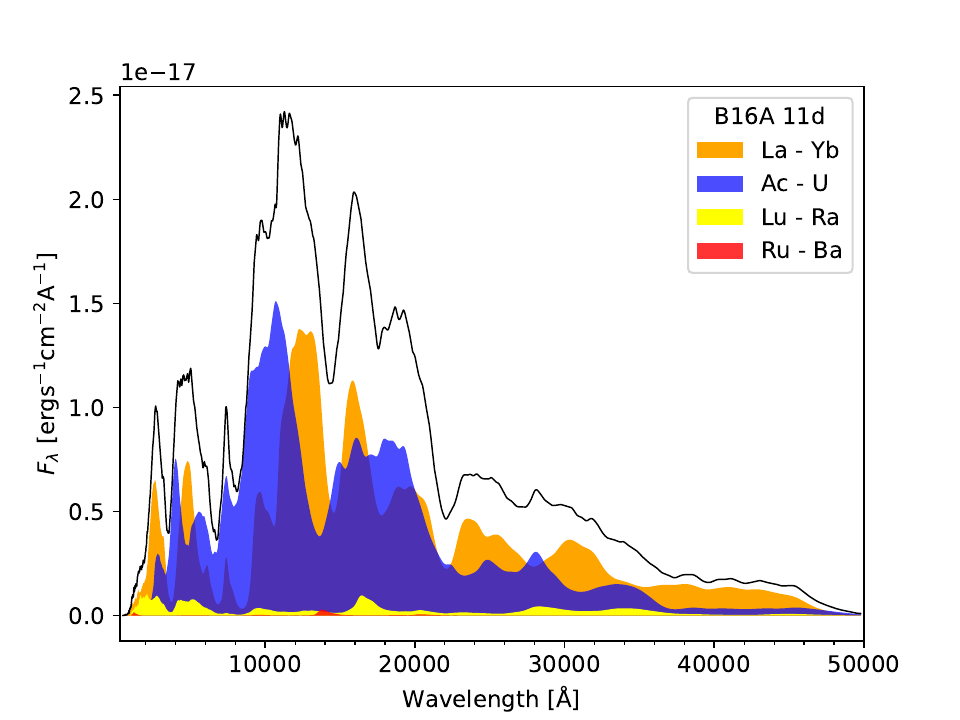}
    \includegraphics[trim={0.4cm 0cm 0.4cm 0.2cm},width=1.0\linewidth]{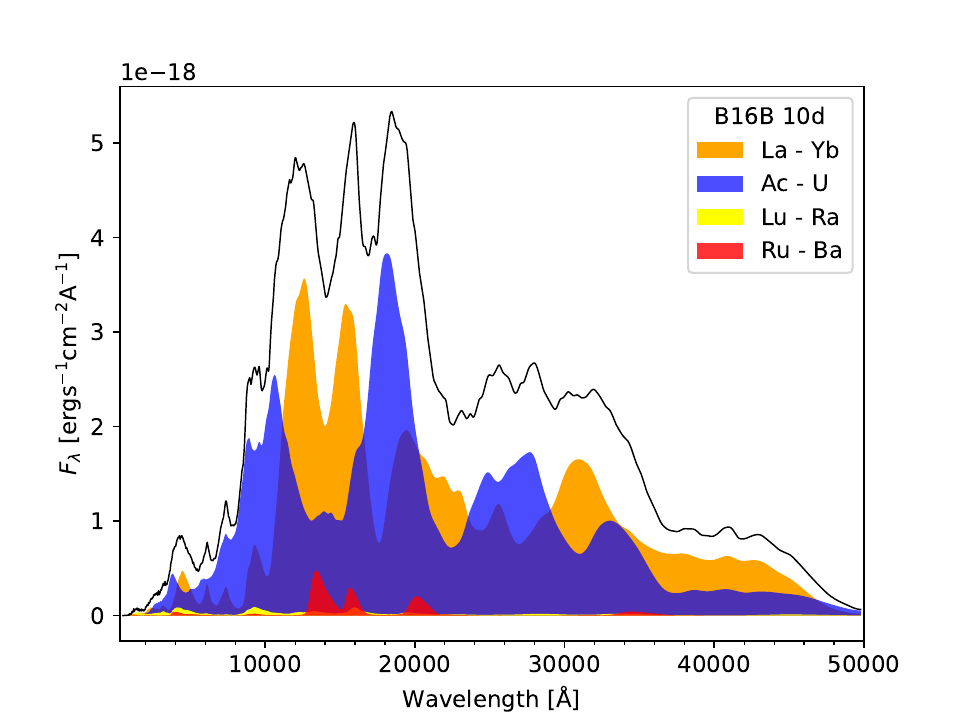}
    \includegraphics[trim={0.4cm 0cm 0.4cm 0.2cm},width=1.0\linewidth]{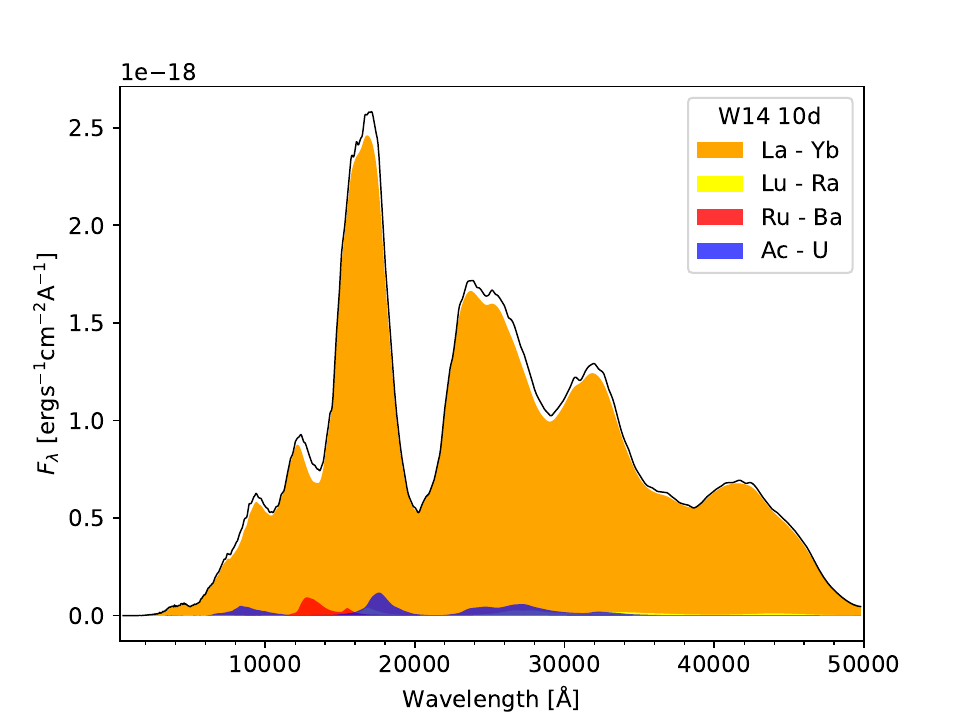}
    \caption{Contribution of element groups to the emergent spectra of the models at 11 days for B16A (top panel) and 10 days for B16B and W14 (middle and bottom panel respectively). Note the different flux levels, where the more energetic models are brighter.}
    \label{fig:species_10d}
\end{figure}

\begin{figure*}
    \centering
    \includegraphics[trim={0.4cm 0cm 0.4cm 0.2cm},width=0.8\linewidth]{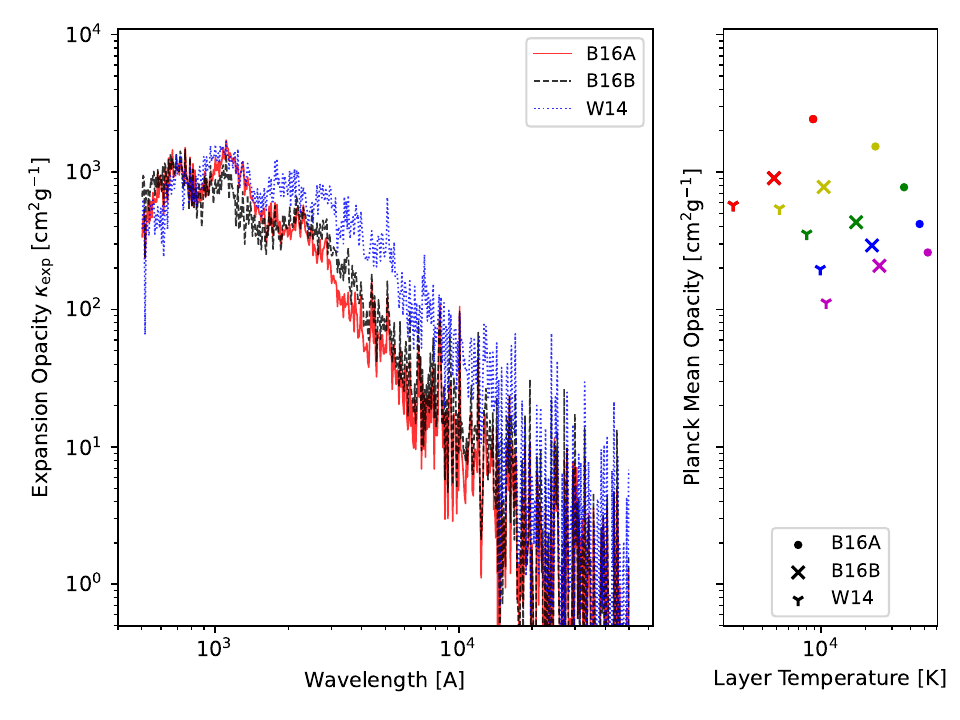}
    \caption{Expansion opacities averaged over zone mass for each model (left panel) and Planck mean opacities for each layer (right panel) at 11 days for B16A, and 10 days for B16B and W14. The color scheme of the markers in the right panel follows the same convention as in Fig. \ref{fig:thermo}, i.e. red corresponds to the innermost layer and magenta to the outermost.}
    \label{fig:exp_opacity}
\end{figure*}

\section{Spectra}
\label{sec:spectral_results}

\subsection{Actinide-rich and actinide-free spectra} 
\label{subsec:spec_compare}

We first compare the emergent spectra of the actinide-rich B16A and B16B models, to the actinide-free spectrum of the W14 model at the earliest epochs studied here. The overall contributions of the various r-process element groups for each model spectra at $\sim 10$ days can be seen in Fig. \ref{fig:species_10d}. There is a striking lack of actinide features in the W14 model, arising from the small mass-fraction of these elements in the composition ($X_{\rm{Ac}}\sim 10^{-3}$, see Table \ref{tab:compositions}). Instead, as the most lanthanide-rich model with $X_{\rm{La}} \sim 0.26$, the emergent spectrum is entirely dominated by lanthanide species, with this dominance remaining throughout all epochs studied here (see Appendix \ref{app:W14_spectra}). The W14 model also has a much lower emergent luminosity, following directly from the lower energy input (see Fig. \ref{fig:dep_compare}). 

It is interesting to note that for the relatively late epochs studied here of $\sim$10 -- 100 days, an actinide mass fraction of $10^{-3}$ is insufficient to produce notable spectral signatures, while a much smaller mass fraction of $10^{-5}$ was found to still produce significant absorption from Th\,\textsc{iii} at very early (1.5 -- 2.5 days) times \citep[][]{Domoto.etal:24}. In that case however, the lanthanide fraction was also considerably smaller than in the W14 model, at $X_{\rm{La}} \sim 6\times10^{-4}$ cf. $X_{\rm{La}} \sim 2\times10^{-2}$. This implies that actinides may still produce significant spectral features even at very low abundances, provided the lanthanide abundance is not vastly superior. These lanthanides may suppress actinide features when sufficiently abundant, as is likely the case in our W14 model. The notably different ejecta conditions between early photospheric epochs, and the later nebular phase studied here, should however be kept in mind when making direct comparisons, as the ejecta conditions are highly different and therefore may potentially lead to equally different emergent spectra.

Following the above argument, it is perhaps not entirely surprising to find that even with relatively small actinide mass fractions of $X_{\rm{Ac}} \sim 0.018 - 0.030$ in the two B16A/B models, we recover significant actinide emission in the overall spectrum. As open f-shell species, actinides have a large energy level count and correspondingly numerous transitions. This allows them to efficiently absorb blue optical photons and produce redder photons through fluorescence cascades, even more effectively than the lanthanides \citep[][]{Flors.etal:23,Fontes.etal:23,Pognan.etal:23}. Several low lying transitions from actinide ions that are no longer open f-shell species, such as Ac \textsc{iii} and U \textsc{iv} described in the next subsection, also produce significant flux by direct emission, as opposed to fluorescence. These factors combined likely lead to the actinides providing anywhere from $\sim$ 40 -- 80 per cent of the emergent model fluxes, despite their small mass fractions. 

The variations between the three models highlight the importance of a proper accounting of effects arising from choices of thermodynamic expansion history, as well as uncertainties in nuclear network inputs, on ejecta composition and emergent KN spectra. Notably, techniques making use of quantities such as expansion opacities and heating-rates parametrized by $Y_e$ in order to broadly infer ejecta properties are particularly susceptible. As is shown here, for a given $Y_e$, composition, radioactive power, and emergent spectra may be entirely different depending on which nuclear inputs are used, which for the low $Y_e$ ejecta considered here, are mostly based on theory as opposed to experimental measurement. This remains true when comparing models that have relatively similar total-heating rates, but distributed over different decay products, such as the B16B and W14 models in this case.

This impact is portrayed in Fig. \ref{fig:exp_opacity}, which shows the different expansion opacities and Planck mean opacities for the three models at the same epochs. These are found to be significantly distinct, following the aforementioned differences. Thus, expansion opacity calculations binned by $Y_e$ that are broadly used in radiative transfer simulations, or in the analysis of observations, must be well aware of the underlying nuclear physics being employed to yield the inputs to the opacity calculations, particularly when dealing with low $Y_e$ ejecta most susceptible to these uncertainties.

\begin{figure*}
    \centering
    \includegraphics[trim={0.5cm 0cm 0.6cm 0.2cm},width=0.49\linewidth]{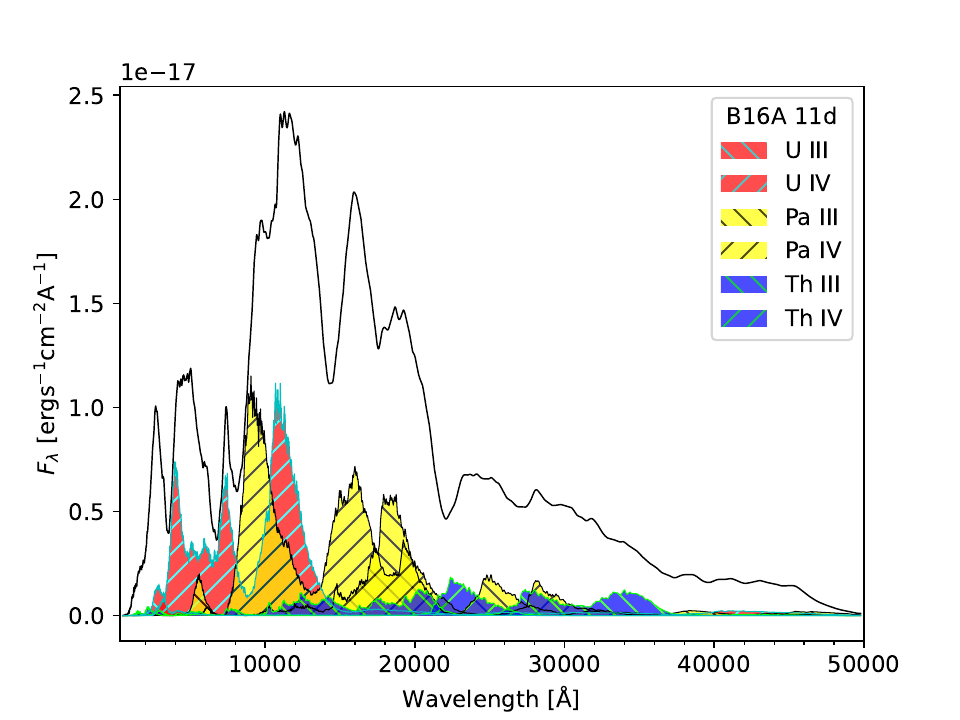}
    \includegraphics[trim={0.5cm 0cm 0.6cm 0.2cm},width=0.49\linewidth]{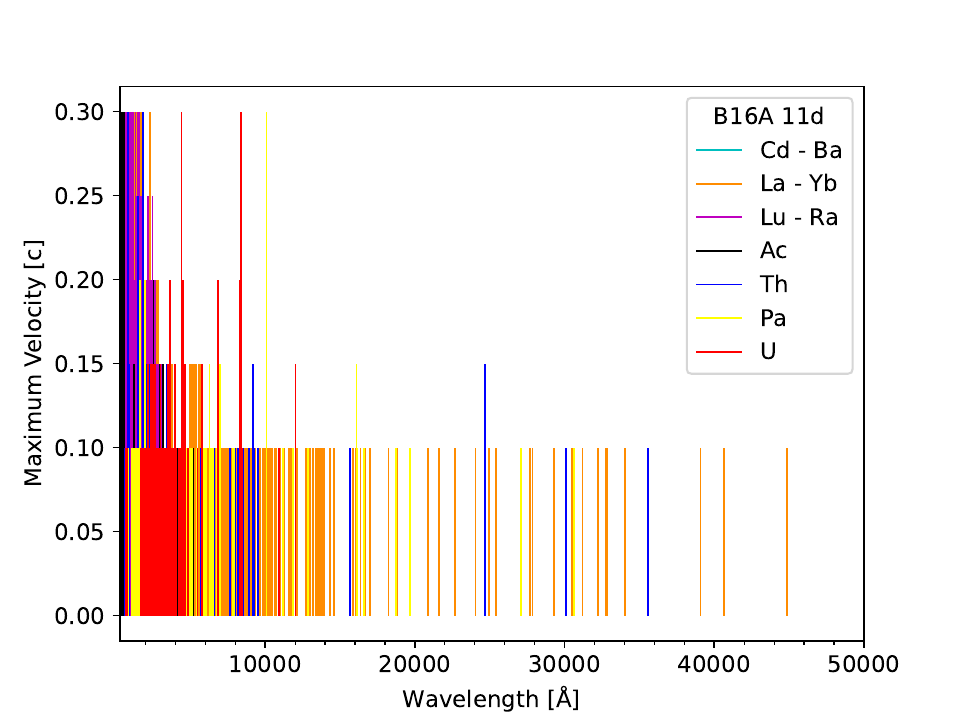}
    \includegraphics[trim={0.5cm 0cm 0.6cm 0.2cm},width=0.49\linewidth]{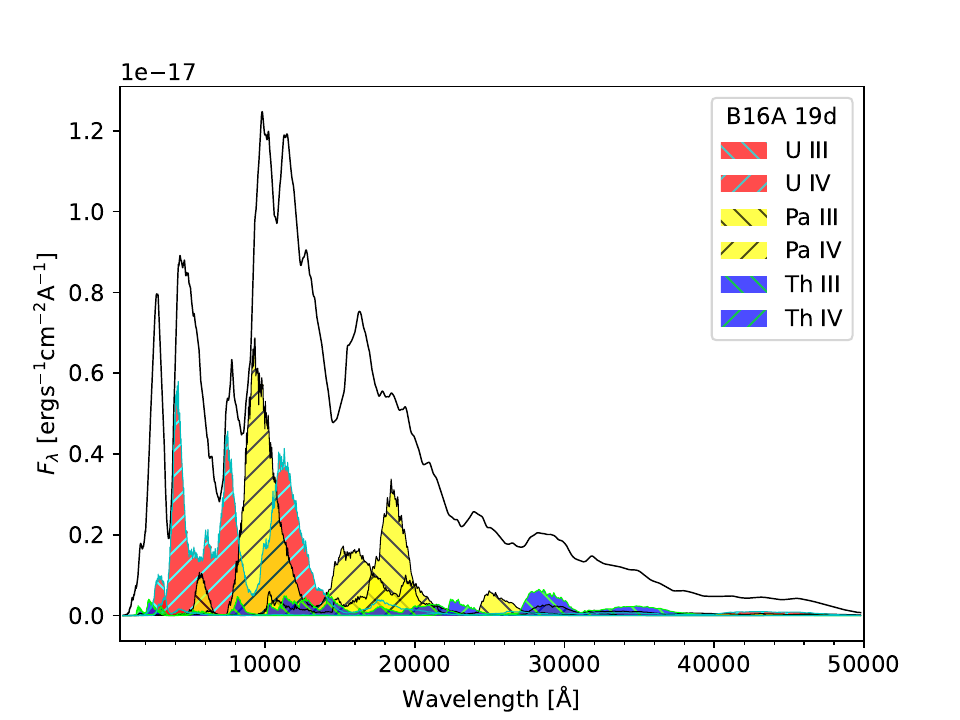}
    \includegraphics[trim={0.5cm 0cm 0.6cm 0.2cm},width=0.49\linewidth]{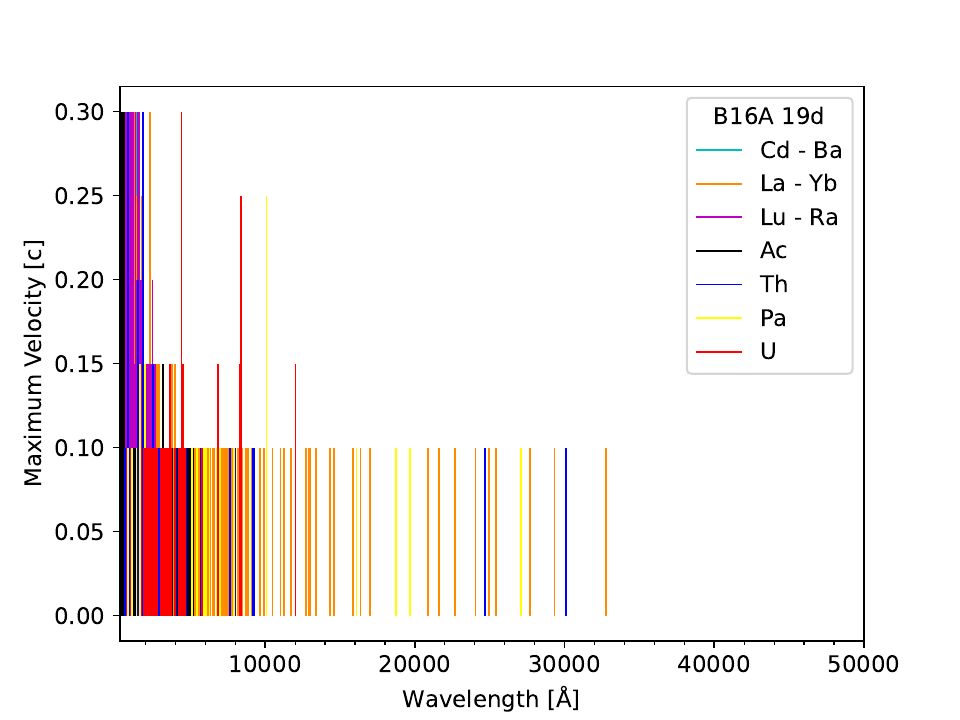}
    \includegraphics[trim={0.5cm 0cm 0.6cm 0.2cm},width=0.49\linewidth]{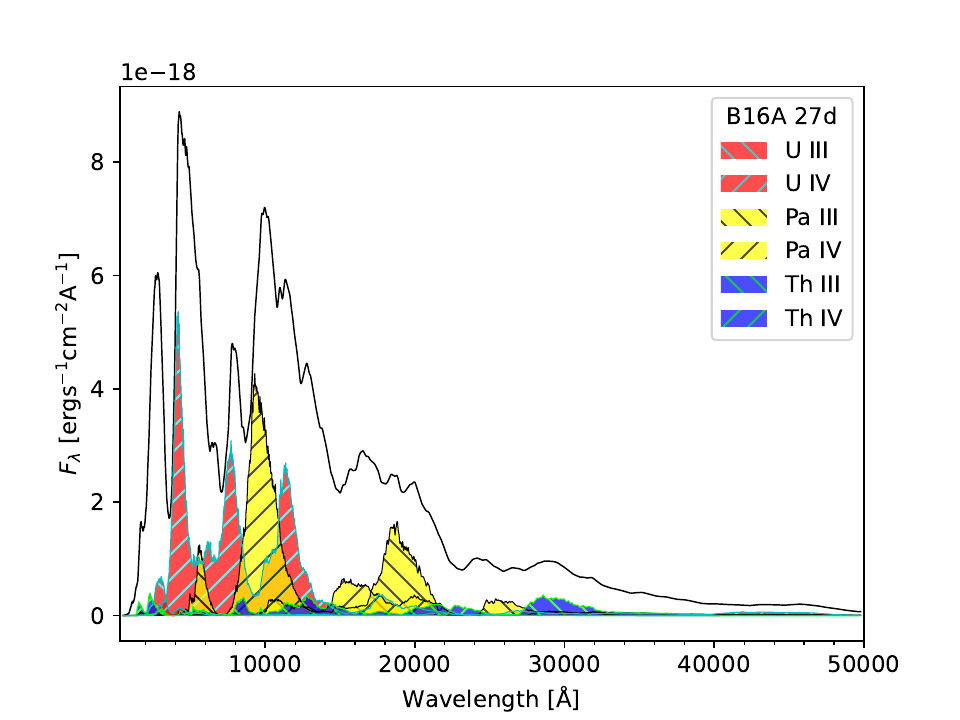}
    \includegraphics[trim={0.5cm 0cm 0.6cm 0.2cm},width=0.49\linewidth]{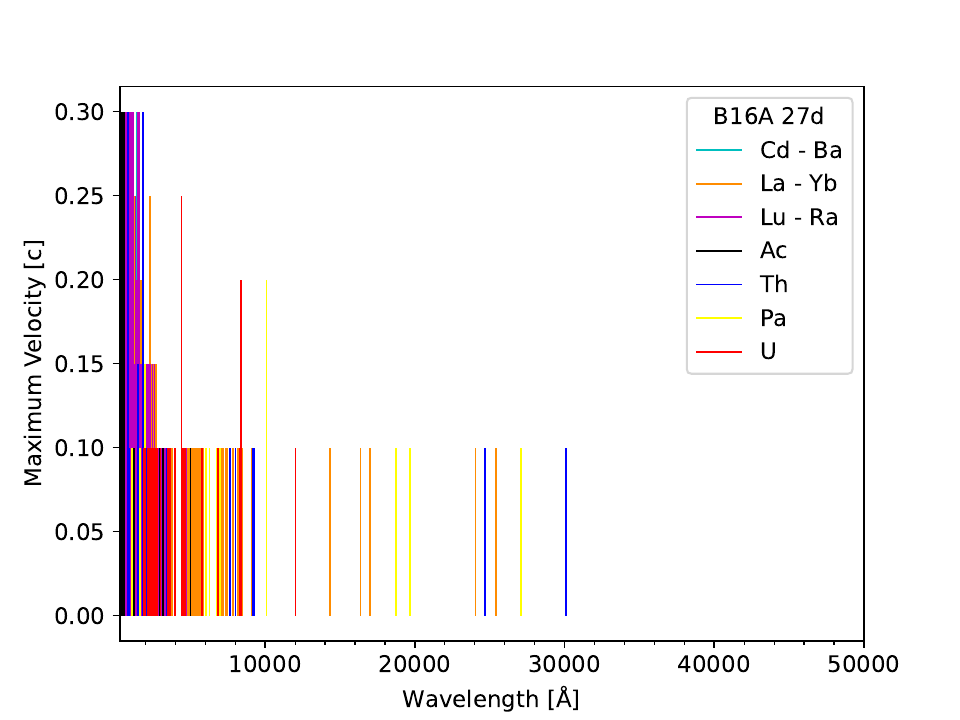}
    \caption{The B16A model spectra with actinide contributions marked out (left-hand panels) and corresponding optically thick lines ($\tau \geq 1$) at rest wavelength (right-hand panels), at 11, 19 and 27 days respectively.}
    \label{fig:B16A_actinides_10_30d}
\end{figure*}

\begin{figure*}
    \centering
    \includegraphics[trim={0.5cm 0cm 0.6cm 0.2cm},width=0.49\linewidth]{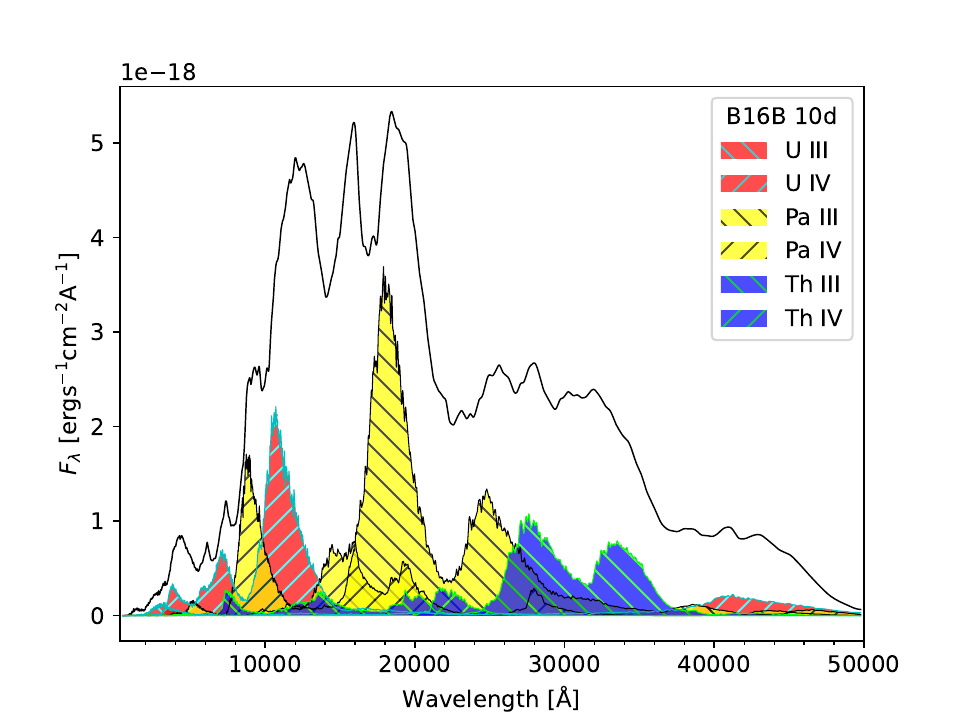}
    \includegraphics[trim={0.5cm 0cm 0.6cm 0.2cm},width=0.49\linewidth]{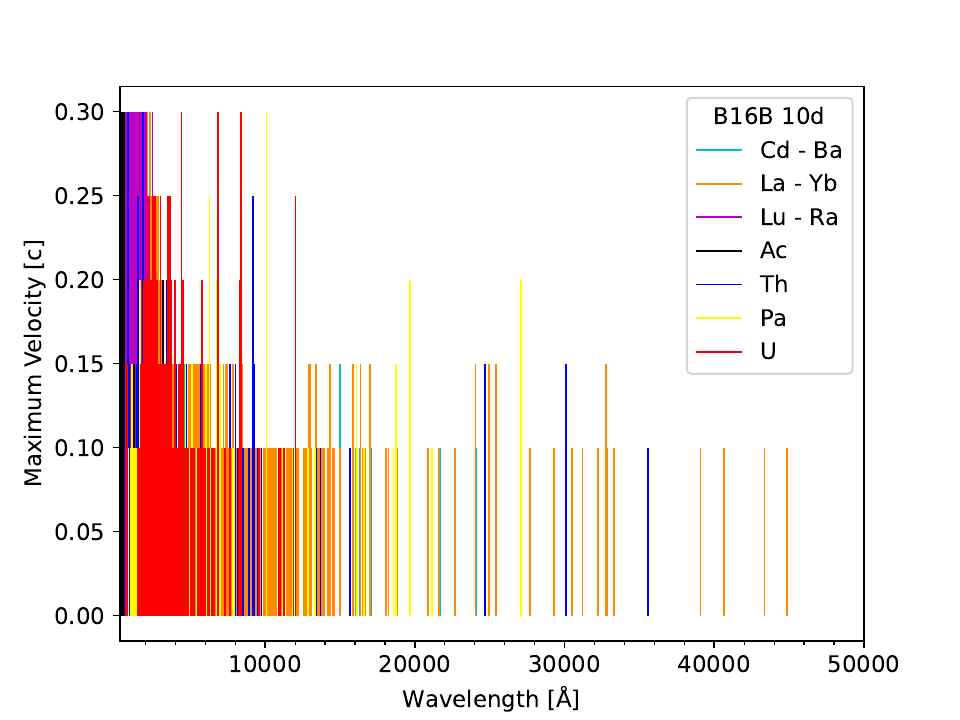}
    \includegraphics[trim={0.5cm 0cm 0.6cm 0.2cm},width=0.49\linewidth]{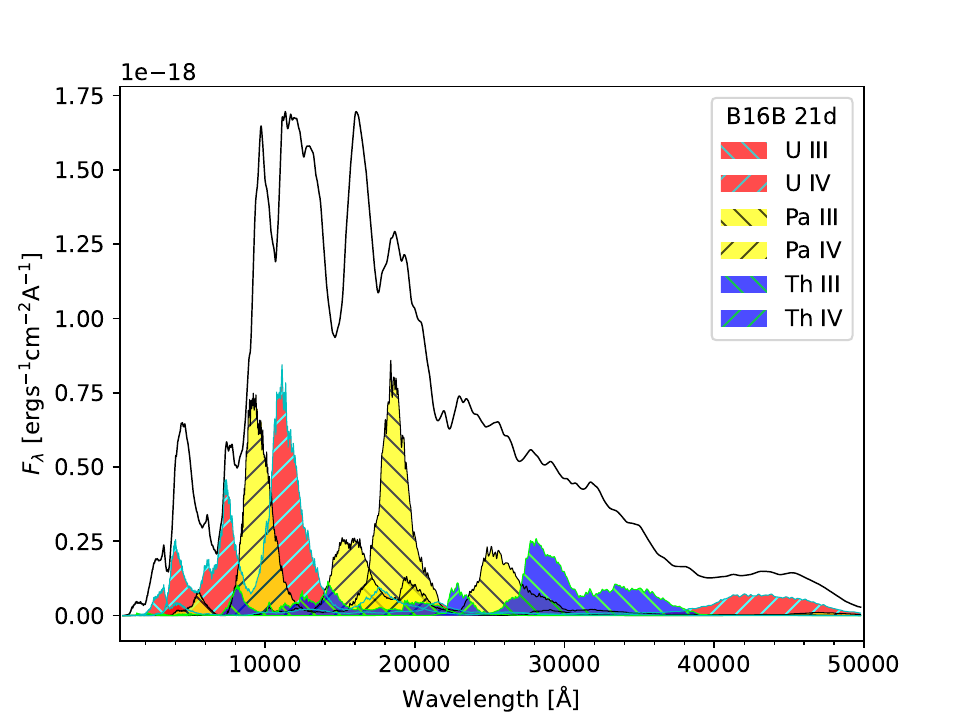}
    \includegraphics[trim={0.5cm 0cm 0.6cm 0.2cm},width=0.49\linewidth]{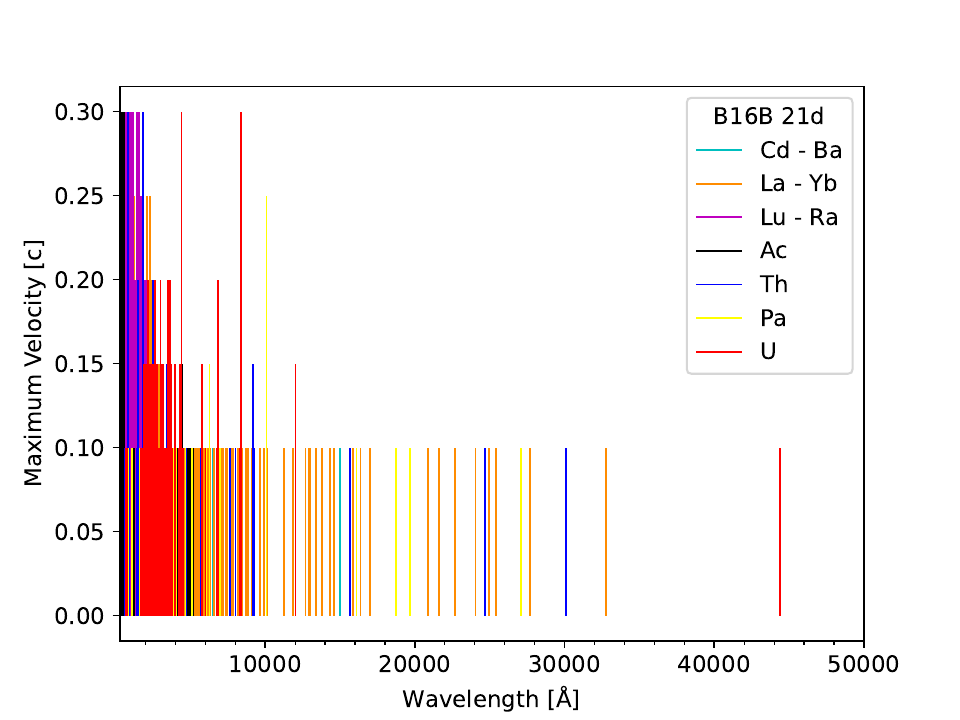}
    \includegraphics[trim={0.5cm 0cm 0.6cm 0.2cm},width=0.49\linewidth]{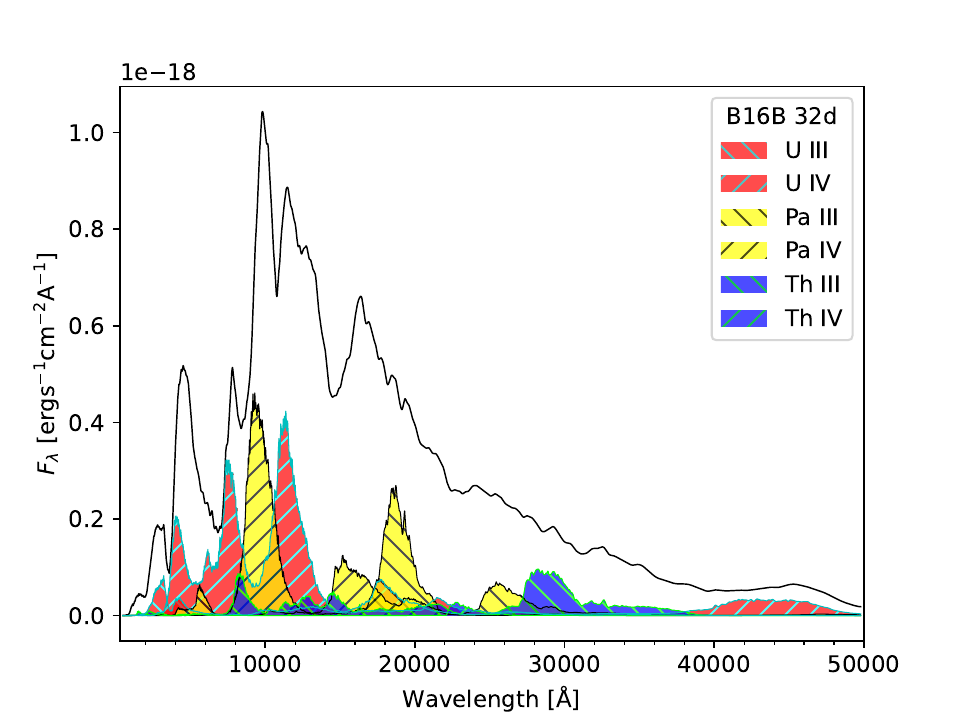}
    \includegraphics[trim={0.5cm 0cm 0.6cm 0.2cm},width=0.49\linewidth]{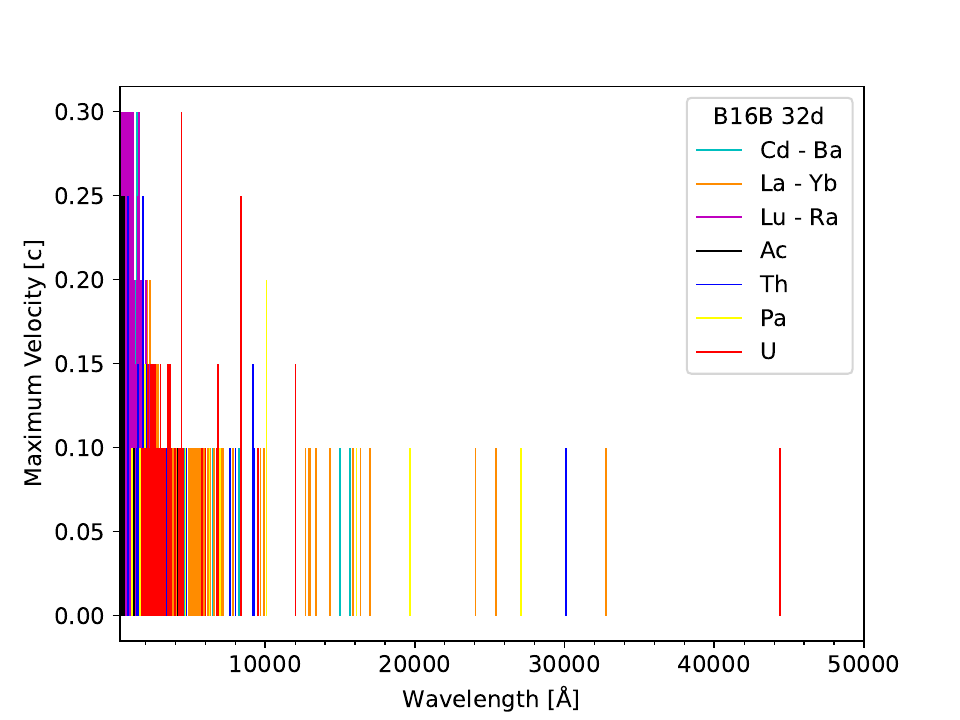}
    \caption{The B16B model spectra with actinide contributions marked out (left-hand panels) and corresponding optically thick lines ($\tau \geq 1$)  at rest wavelength (right-hand panels), at 11, 21 and 32 days respectively.}
    \label{fig:B16B_actinides_10_30d}
\end{figure*}

\subsection{Spectral features and evolution of the actinide-rich models}
\label{subsec:spec_actinides}

\begin{figure*}
    \centering
    \includegraphics[trim={0.5cm 0cm 0.6cm 0.2cm},width=0.49\linewidth]{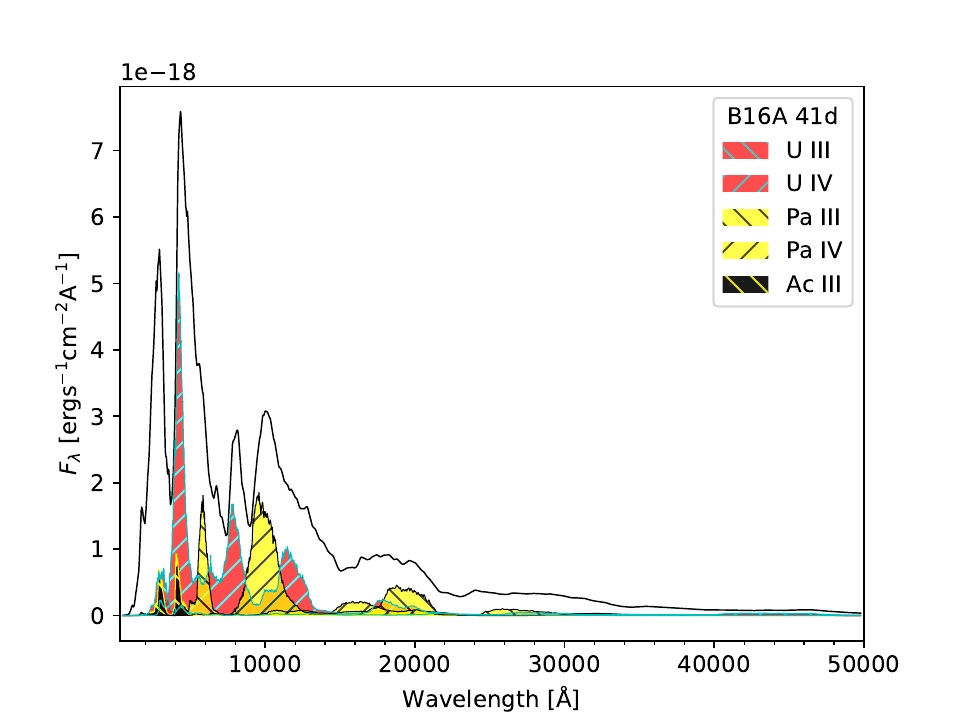}
    \includegraphics[trim={0.5cm 0cm 0.6cm 0.2cm},width=0.49\linewidth]{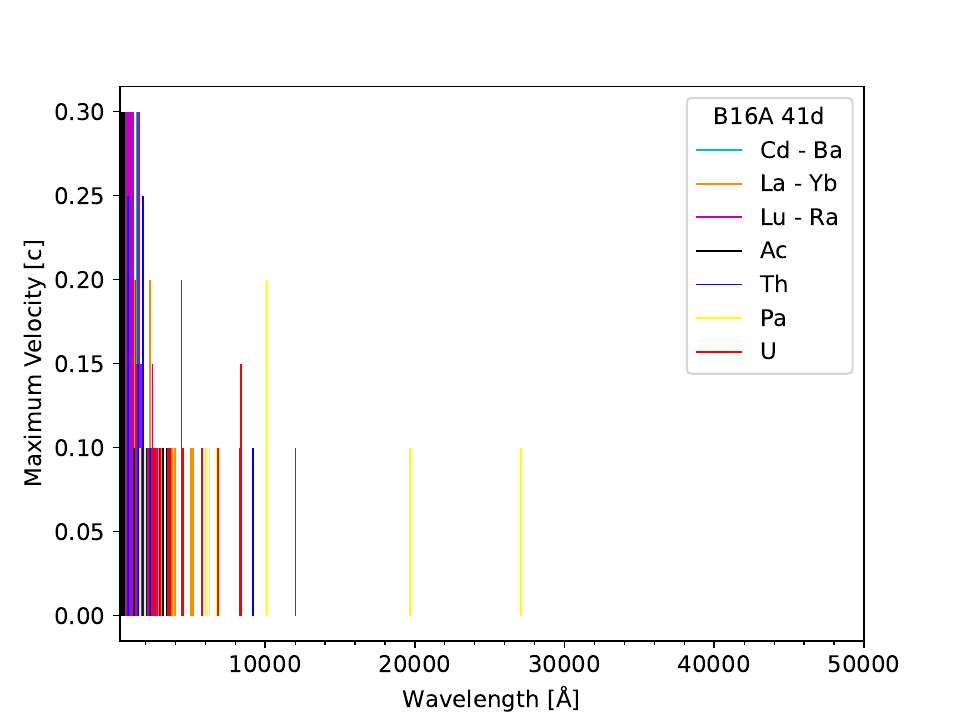}
    \includegraphics[trim={0.5cm 0cm 0.6cm 0.2cm},width=0.49\linewidth]{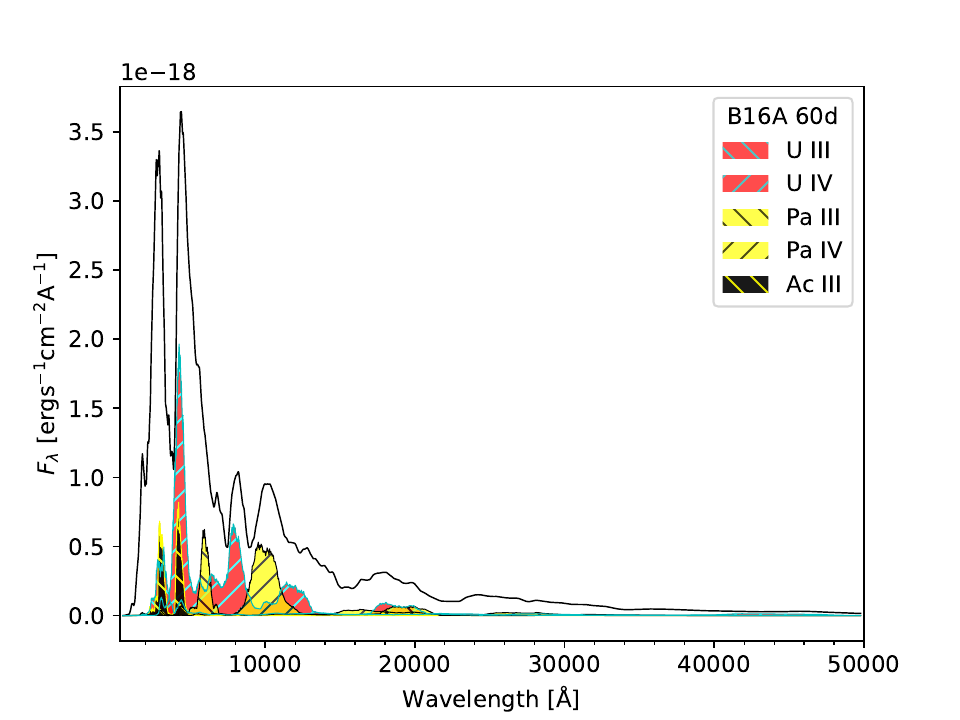}
    \includegraphics[trim={0.5cm 0cm 0.6cm 0.2cm},width=0.49\linewidth]{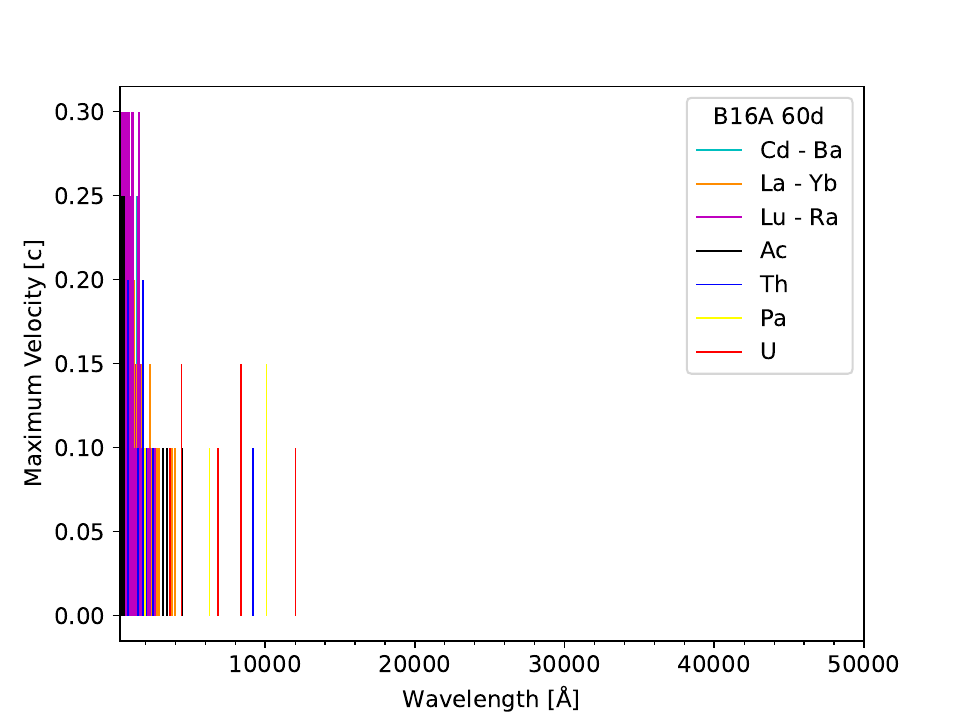}
    \includegraphics[trim={0.5cm 0cm 0.6cm 0.2cm},width=0.49\linewidth]{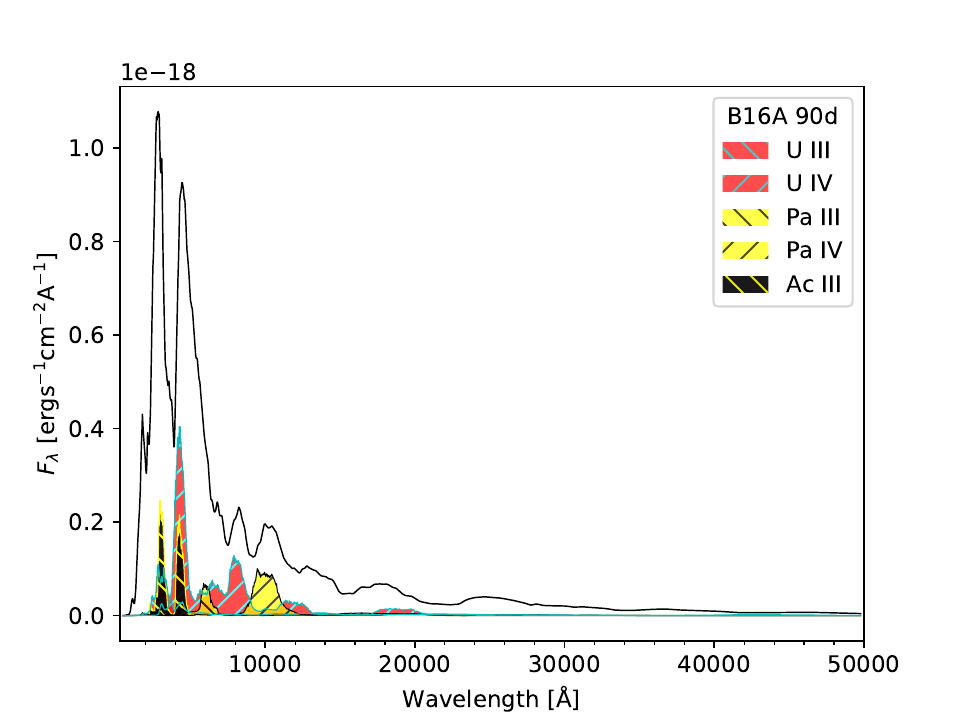}
    \includegraphics[trim={0.5cm 0cm 0.6cm 0.2cm},width=0.49\linewidth]{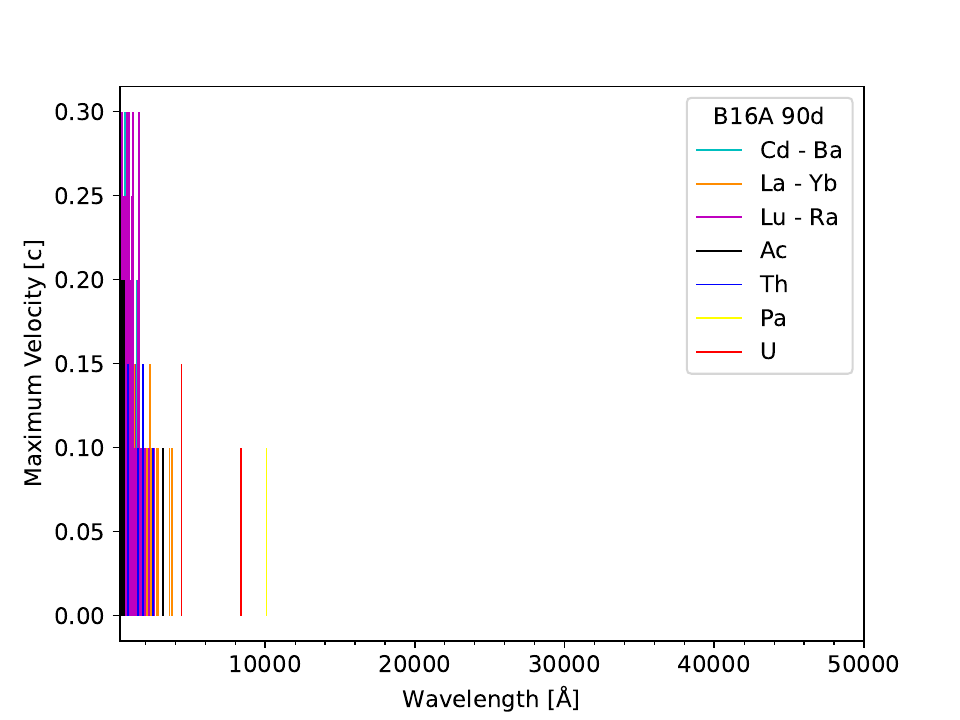}
    \caption{The B16A model spectra with actinide contributions marked out (left-hand panels) and corresponding optically thick lines ($\tau \geq 1$)  at rest wavelength (right-hand panels), at 41, 60 and 90 days respectively.}
    \label{fig:B16A_actinides_40_90d}
\end{figure*}

\begin{figure*}
    \centering
    \includegraphics[trim={0.5cm 0cm 0.6cm 0.2cm},width=0.49\linewidth]{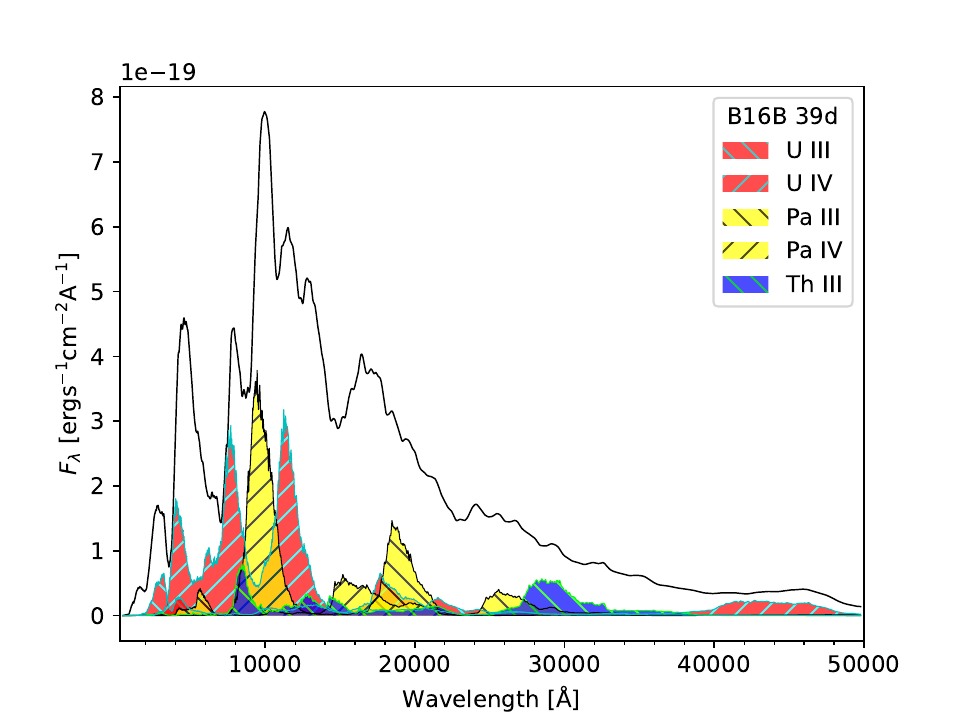}
    \includegraphics[trim={0.5cm 0cm 0.6cm 0.2cm},width=0.49\linewidth]{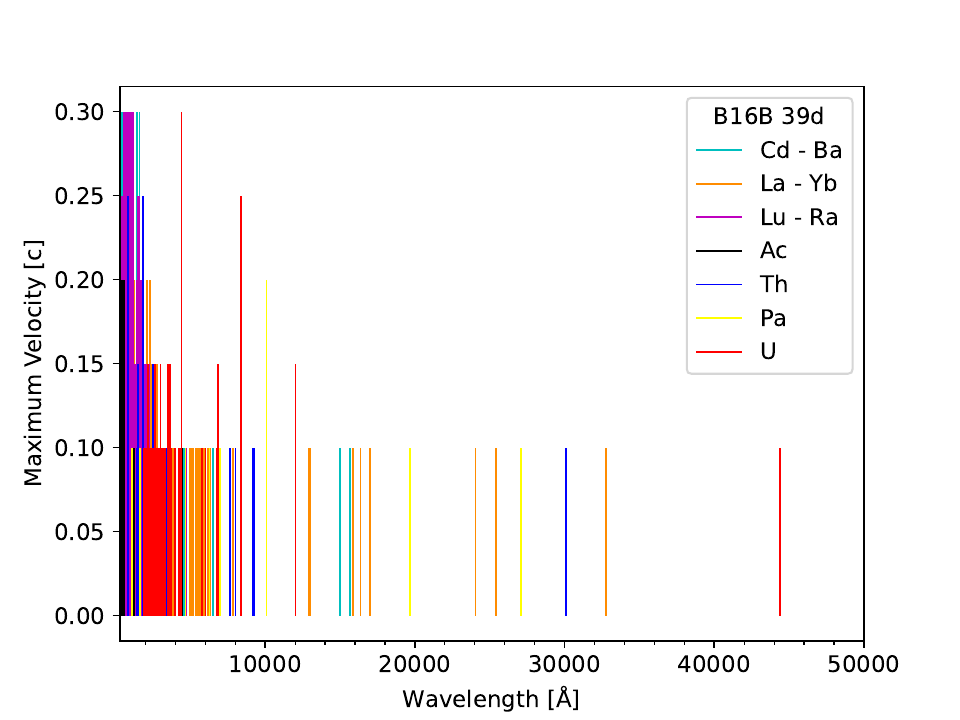}
    \includegraphics[trim={0.5cm 0cm 0.6cm 0.2cm},width=0.49\linewidth]{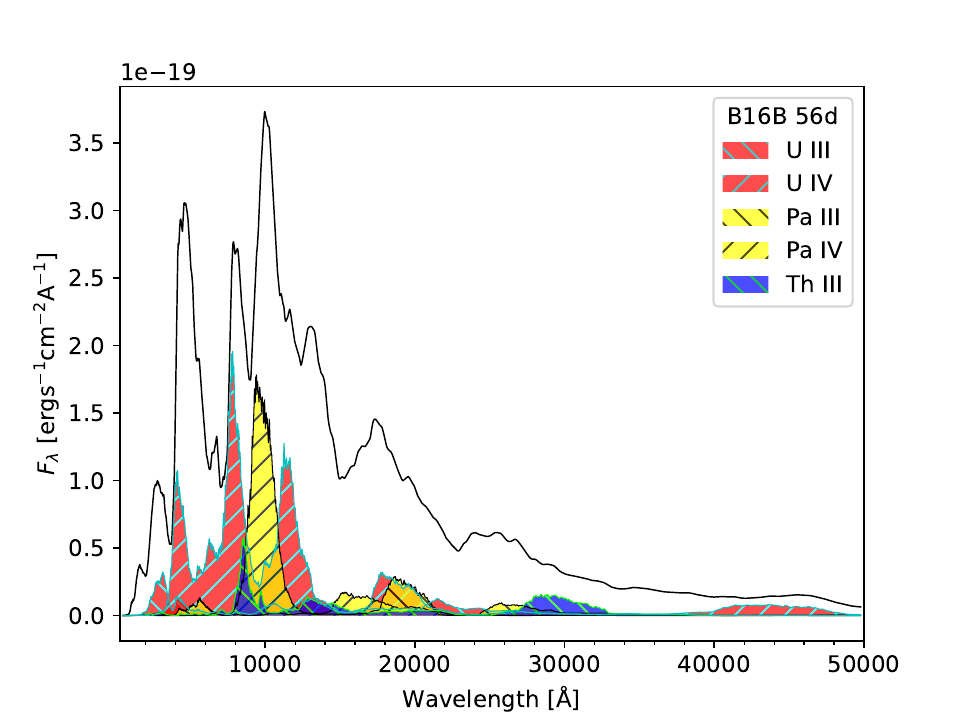}
    \includegraphics[trim={0.5cm 0cm 0.6cm 0.2cm},width=0.49\linewidth]{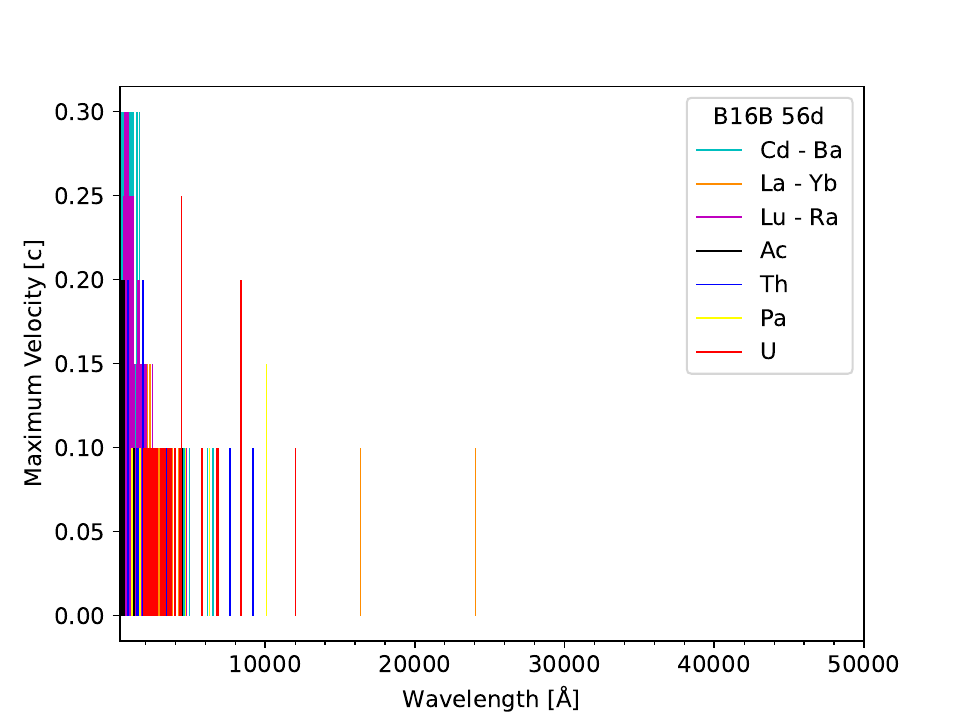}
    \includegraphics[trim={0.5cm 0cm 0.6cm 0.2cm},width=0.49\linewidth]{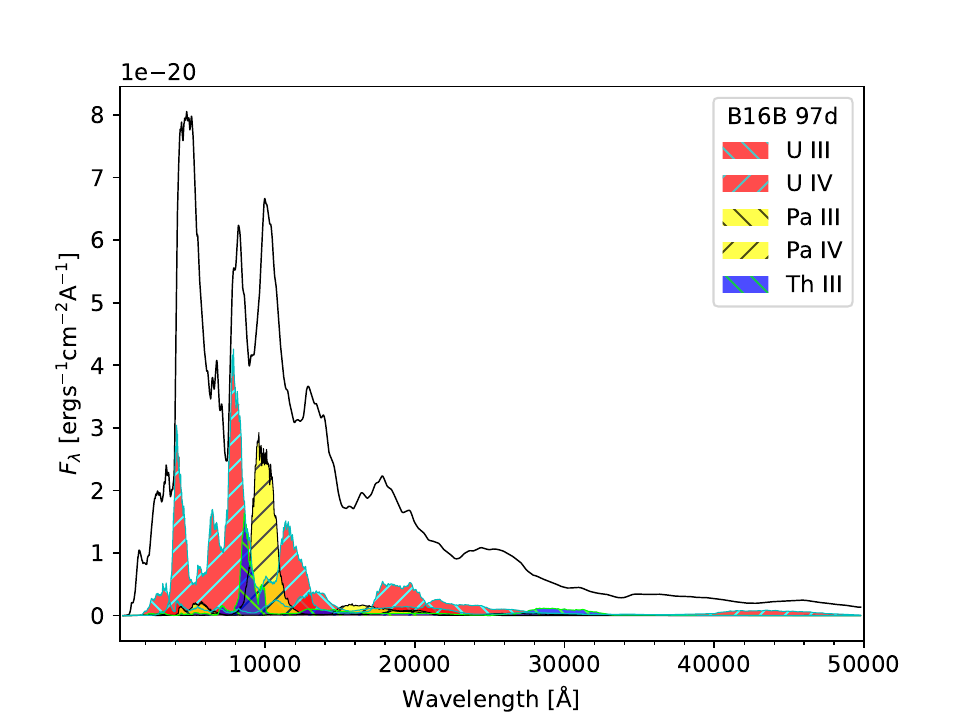}
    \includegraphics[trim={0.5cm 0cm 0.6cm 0.2cm},width=0.49\linewidth]{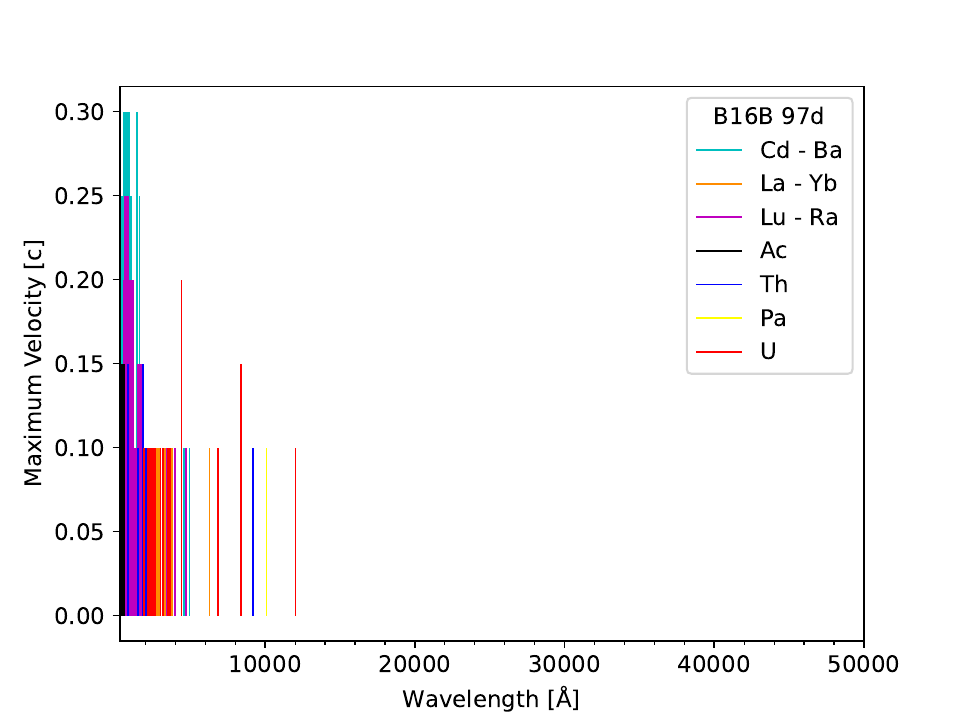}
    \caption{The B16B model spectra with actinide contributions marked out (right-hand panels) and corresponding optically thick lines ($\tau \geq 1$)  at rest wavelength (left-hand panels), at 39, 56 and 97 days respectively.}
    \label{fig:B16B_actinides_40_90d}
\end{figure*}

The contribution of individual actinide species, as well as the optically thick lines present in the models, at selected epochs can be viewed in Figs.~\ref{fig:B16A_actinides_10_30d} to~\ref{fig:B16B_actinides_40_90d}. n these figures, the model spectra are shown in the left-hand panels, and the corresponding optically thick lines in the right-hand panels. The line plots show optically thick ($\tau \geq 1$) transitions in the 5 model zones from velocity coordinates $v_{\rm{inner}} = 0.05$c to $v_{\rm{outer}} = 0.3$c, in steps of 0.05c. The height of the markers in the plot shows to which point in the ejecta the transition is optically thick e.g. if the marker reaches 0.2c, the transition is optically thick out until the third ejecta layer. In the analysis of the spectra below, we use the optically thick line plots to understand where absorption troughs in the spectra originate from. Since the W14 model does not exhibit marked actinide features, we focus here on the B16A and B16B models, while the spectral evolution of the W14 model can be viewed in Appendix~\ref{app:W14_spectra}. 

\subsubsection{Early times: 10 - 30 days}

Considering first Fig. \ref{fig:B16A_actinides_10_30d}, we have the B16A model spectra and optically thick lines from 11 to 27d. Looking at the 11 days epoch in the top panels, the most active actinide species in spectral formation are found to be doubly and triply ionized $_{91}$Pa and $_{92}$U, with some weaker contributions from $_{90}$Th. The main actinide-dominated features here are the $_{92}$U P-Cygni like features at $\sim 4000$ and $\sim 8000 \ang$ respectively, and the blended $_{91}$Pa and $_{92}$U feature at 1 micron. The double feature from Pa\,\textsc{iii} and Pa\,\textsc{iv} at 1.7 micron contributes significantly to the emergent spectral shape at this wavelength, but does not entirely dominate, being heavily blended with lanthanide contributions (see Fig. \ref{fig:species_10d}). The U\,\textsc{iv} transitions, as well as the Pa\,\textsc{iv} transition at 1 micron, are found to be strong ($A \gtrsim 10^{5} \, \rm{s^{-1}}$) allowed transitions to the ground state. Conversely, the blended $_{91}$Pa feature at 1.7 micron results from the contribution of many diverse allowed transitions with smaller radiative transition rates of $A \lesssim 10^{4} \, \rm{s^{-1}}$, typically seen in fluorescence cascades. 

From the right-hand side of the top panels, we see that $_{91}$Pa and $_{92}$U have a few lines that are optically thick throughout the whole ejecta, and which line up with P-Cygni like features below 1 micron. We also note that the majority of the ejecta are optically thin beyond the innermost layer from about 1.8 micron and redwards, with the innermost layer only having a few optically thick lanthanide, $_{90}$Th, and $_{91}$Pa lines. At shorter wavelengths $\lambda \lesssim 6000 \ang$, we find a blend of many diverse optically thick lines, with important contributions from both the lanthanide species, as well as third r-process peak species. 

Looking now at the middle and bottom panels of Fig. \ref{fig:B16A_actinides_10_30d}, we can see how the spectra and actinide signatures evolve with time in the B16A model. The spectral energy distribution (SED) follows the general evolution first seen in \citet{Pognan.etal:23}, with the spectra gradually becoming more blue with time. From the right-hand panels, we find that the ejecta begins to become entirely optically thin at redder wavelengths past $\sim 3$ micron, with the outermost layer also becoming optically thin below 1 micron down to 2000 $\ang$. The aforementioned actinide-dominated spectral features evolve in relatively minor ways in this 10 - 30 day timespan, with red features generally fading more than the bluer features.

Fig. \ref{fig:B16B_actinides_10_30d} shows the same plots as those described above, but for the B16B model. In the top panels, we see that the B16B model has actinide features dominated by the same species and ionization states as the B16A model, albeit with different relative contributions. For instance, more emission from Th\,\textsc{iii} at 2.8 and 3.4 micron features is found, which are low lying allowed E1 transitions with $\lambda_0 = 30096\, \ang$, $\rm{A} \sim 10^4\, \rm{s^{-1}}$, and  $\lambda_0 = 35551\, \ang$, $\rm{A} \sim 10^{3} \, \rm{s^{-1}}$ respectively. In terms of optically thick lines, the B16A/B models are broadly similar, with the B16B model having somewhat more optical depth at redder wavelengths, and further out into the ejecta.

Considering the evolution of the B16B model by looking at the middle and bottom panels of Fig. \ref{fig:B16B_actinides_10_30d}, we find that U\,\textsc{iv} begins to dominate the spectrum past 4~$\mu$m. The contribution of the Pa\,\textsc{iii} emission at 1.8 micron is greatly reduced with time, which is somewhat opposite of the behaviour seen in the B16A model, where the feature becomes dominant at that wavelength. Similarly to the B16A model, the B16B model becomes almost entirely optically thin past 3 micron, with a single optically thick U\,\textsc{iv} line in the innermost layer at 4.4 micron, the main contributor to U\,\textsc{iv} emission at that same wavelength. This transition is found to be an E1 dipole transition with a radiative transition rate of $A \sim 10^{3} \, \rm{s^{-1}}$.

\subsubsection{Late times: 40 - 100 days}

The later evolution of actinide features up to $\sim 90$ days are shown in Figs. \ref{fig:B16A_actinides_40_90d} and \ref{fig:B16B_actinides_40_90d} for the B16A and B16B models respectively. It is worth initially drawing attention to how slow the spectral evolution becomes at these late times. We note that the spectral shapes remain remarkably similar, aside from dropping flux levels and minor feature changes. The different model temperatures at these later epochs have a significant effect on the emergent spectra, as we see that the B16A model produces notably bluer features, with a smooth, featureless NIR continuum. In both models, we see that the entire ejecta becomes optically thin past 1 micron at later times, around 40~days for the B16A model, and closer to 60~days for the B16B model. Conversely, significant optical depth remains at the bluest wavelengths $\lambda \lesssim 4000$~$\ang$ in both models, for the entire duration of the timespan studied here. 

\begin{table*}
    \caption{Emitted flux of W\,\textsc{iii} at 4.75 and 5.22 micron compared to the total emitted flux of the models, in units of $\rm{erg\,s^{-1}\,cm^{-2}\,\ang^{-1}}$. The B16A model is at 41 days, and the B16B/W14 models at 47 days.}
    \label{tab:WIII_emission}
    \centering
    \begin{threeparttable}
    \bgroup
    \def\arraystretch{1.2}
    \setlength\tabcolsep{0.12cm}
     \begin{tabular}{cccccccccccc}
         \hline \hline 
       & \multicolumn{3}{c}{B16A} & & \multicolumn{3}{c}{B16B} & & \multicolumn{3}{c}{W14} \\ \cline{2-4} \cline{6-8} \cline{10-12}
      $\lambda$ [$\mu$m]  & $F_{\lambda}^{\rm{tot}}$ &  $F_{\lambda}^{\rm{W\,\textsc{iii}}}$ & $F_{\lambda}^{\rm{W\,\textsc{iii}}}/F_{\lambda}^{\rm{tot}}$  & &  $F_{\lambda}^{\rm{tot}}$ &  $F_{\lambda}^{\rm{W\,\textsc{iii}}}$ & $F_{\lambda}^{\rm{W\,\textsc{iii}}}/F_{\lambda}^{\rm{tot}}$ & & $F_{\lambda}^{\rm{tot}}$ &  $F_{\lambda}^{\rm{W\,\textsc{iii}}}$ &  $F_{\lambda}^{\rm{W\,\textsc{iii}}}/F_{\lambda}^{\rm{tot}}$ \\
       \hline 
    4.75 & $8.448\times10^{-20}$ & $7.296\times10^{-21}$ & 0.086 & & $2.488\times10^{-20}$ & $4.270\times10^{-22}$ & 0.017 & & $3.562\times10^{-20}$ & $3.059\times10^{-21}$ & 0.086 \\
    5.22 &  $4.534\times10^{-20}$ & $7.912\times10^{-21}$ & 0.175 & & $1.723\times10^{-20}$ & $5.020\times10^{-22}$ & 0.029 & & $3.257\times10^{-20}$ & $3.828\times10^{-21}$ & 0.117 \\
    \hline \hline
    \end{tabular}
    \egroup
\end{threeparttable}
\end{table*}

The differences in emergent features between the B16A and B16B models are likely due more to the different temperature and ionization structure solutions, as opposed to differing abundances of actinide species. Notably, the B16B model is cooler and less ionized, such that features arising from doubly ionized species are typically stronger. Conversely, the B16A model has more important contributions from triply ionized species. This is well visualized by the relative Pa\,\textsc{iii} and Pa\,\textsc{iv} contributions to the emission at $\sim 1.8$~$\mu$m, where we see roughly equal contributions from these two species in the B16A model, but domination of Pa\,\textsc{iii} in the B16B model. 

At later times $\gtrsim 60$ days, the B16A model develops two Ac\,\textsc{iii} features at $\sim 3200, 4400 \ang$ respectively, which blend with the U\,\textsc{iii} and U\,\textsc{iv} features, as well as emission from Tm\,\textsc{iii} and Eu\,\textsc{iii}, and Sm\,\textsc{iii} respectively. However, the $4400~\ang$ emission from Ac\,\textsc{iii} provides approximately half of the actinide flux within the $\sim 5000 \ang$ emission feature. These Ac\,\textsc{iii} emission features are worthy of note, as the atomic data underlying them have wavelengths accurate to within 2 per cent, and A-values within a factor of 2, of data available from NIST. As such, these features are robust predictions that may remain important even if the other features in that area, of which the accuracy is difficult to verify, are corrected to other wavelengths or flux levels. However, it is difficult to predict whether they will become more or less dominant in that regime if the calibration of other atomic species yields a greater number of important transitions in the same wavelength range, though they are expected to produce strong optical features also at early times \citep[][]{Domoto.etal:24}.

\subsubsection{Spectrally significant actinide features}

Of the various actinide features described above, some have a notable impact on the overall spectral shape, which may be detectable in observations. As described in Section \ref{sec:models}, the atomic data underlying the emergent spectra come from theoretical calculations using the \textsc{fac} software, and are not calibrated to existing data, where such data exist. As such, it is important to verify whether the transitions giving rise to important spectral features are accurate and realistic when possible. Unfortunately, no existing literature data for the energy levels of Pa\,\textsc{iii} and Pa\,\textsc{iv} were found, and so we cannot verify the accuracy of the prominent $_{91}$Pa features in our models. 

The Th\,\textsc{iii} emission in our models may be checked by comparing our energy level structure to data available from NIST. In doing so, we find that the wavelengths of our prominent Th\,\textsc{iii} transitions are quite inaccurate, e.g. the emission at $\lambda_0 = 3.55 \mu$m is in reality found to be at $13.0 \mu$m, while the $\lambda_0 = 3.01 \mu$m emission is at $22.3 \mu$m. By also considering the change in radiative transition rate (A-value) arising from shifted energy levels, $A \propto \left(\Delta E_{\rm{NIST}}/\Delta E_{\rm{FAC}} \right)^3$ for dipole transitions \citep[see e.g.][]{Mulholland.etal:24}, we find A-values for the transitions at their correct wavelengths of $A_{13.0 \mu m} \sim 1.7 \times 10^2 \, \rm{s^{-1}}$ and $A_{22.3 \mu m} \sim 2.9 \times 10^1 \, \rm{s^{-1}}$ respectively. While this does mean that the Th\,\textsc{iii} emission in our model spectra cannot be taken directly as potentially observable actinide signatures, these results suggest that Th \textsc{iii} may produce significant mid-IR emission, should the temperature and ionization structure of the ejecta be comparable to those of the B16B model found here. The $22.3~\mu$m transition is particularly interesting, as it corresponds to the $n = 3 \rightarrow 2 $ transition from the lowest lying $\rm{5f^16d^1}$ state to the lowest lying $\rm{6d^2}$ state, and may therefore be a good candidate for late-time, nebular phase observations with \textit{JWST}, given the relatively high A-value and low excitation energies of the involved levels. It would also be interesting to verify if absorption from Th\,\textsc{iii} at 1.8 micron \citep[][]{Domoto.etal:24} persists into the nebular phase, though it is probable that the lines responsible for this absorption in the photospheric phase become optically thin at later times.

The U\,\textsc{iv} emission at 4.4 micron in the B16B model, which dominates flux contribution there past $\sim 30$ days is somewhat reminiscent of the 4.5 micron \textit{Spitzer} detection at 43 and 74 days for AT2017gfo \citep{Kasliwal.etal:22}, which has been tentatively attributed to either Se\,\textsc{iii} or W\,\textsc{iii} by also taking into account the lack of detection at 3.6 micron \citep{Hotokezaka.etal:22}. Looking at Table \ref{tab:compositions}, we see that our models include $_{74}$W, and in significant abundances for the B16A and W14 models. The relevant transitions of W\,\textsc{iii} for 4.5 micron emission are the $n = 3 \rightarrow 2$ and $n= 2 \rightarrow 1$ forbidden transitions with $\lambda = 4.54 \mu$m, $A = 0.61\,\rm{s^{-1}}$ and $\lambda = 4.43 \mu$m, $A = 0.62\,\rm{s^{-1}}$ respectively, following data available from the National Institute of Standards and Technology (NIST) Atomic Spectral Database \citep[][]{NIST_ASD} and \citet{Hotokezaka.etal:22}. The \textsc{fac} data for W\,\textsc{iii} have not been calibrated for this study however, and have these transitions at $\lambda = 4.75 \mu$m, $A = 0.46\,\rm{s^{-1}}$ and $5.22\mu$m, $A = 0.31\,\rm{s^{-1}}$ respectively. The former lies within the spectral range, and therefore may be expected to produce noticeable emission in the emergent spectra, which, however, does not appear to be the case.

This may be so for a variety of reasons. For forbidden transitions to be prominent, we require a balance between low ejecta density for inefficient collisions, but also a degree of ionization that favours W\,\textsc{iii} over W\,\textsc{iv}, i.e. density cannot be so low that recombinations are entirely inefficient. We also need a significant abundance of the species, by elemental composition of the ejecta and ion fraction, as well as the collisional de-excitation time-scale to be longer than the radiative transition lifetime, which is found to be on the order of $\sim 5$\,s in our data. Considering first the B16A model at 41 days, we find that W\,\textsc{iii} is most abundant in the innermost layer, but with a low ion fraction of only $\sim 0.07$, while the subsequent outer layers completely favour W\,\textsc{iv}. Conversely, the B16B model at 47 days has a larger W\,\textsc{iii} ion fraction of $\sim 0.27$, but a much lower $_{74}$W abundance in the model composition to begin with (see Table \ref{tab:compositions}).  As such, the W\,\textsc{iii} emission in both these models may be intrinsically quite low in comparison to other species. 

\begin{table*}
    \caption{The U\,\textsc{iv} levels and associated transitions of interest in our model. The upper $\rm{5f^2 6d^1}$ levels are compared to those in \citet{Seijo.etal:03}, while the lower $\rm{5f^{3}}$ levels to those from \citet{Carnall.Crosswhite:85,Andres.etal:96} (experimental measurements) and \citet{ruiperez.etal:2007,Roy.Prasad:20} (theoretical calculations). }
    \label{tab:UIV_transitions}
    \centering
    \begin{threeparttable}
    \bgroup
    \def\arraystretch{1.2}
    \setlength\tabcolsep{0.4cm}
     \begin{tabular}{ccccccc}
         \hline \hline 
     \multicolumn{2}{c}{Upper level ($\rm{5f^2\,6d^1}$)} &  \multicolumn{2}{c}{Lower level ($\rm{5f^3}$)} & \multicolumn{3}{c}{Wavelength (\ang)} \\ \cline{1-2} \cline{3-4} \cline{5-7}
      $jj$-scheme $J^\pi_n$  & $LS$ main components\tnote{a}  &  $jj$-scheme $J^\pi_n$  & $LS$ main components \tnote{b}  &  FAC  & Exp.\tnote{c} & \textit{Ab initio}\tnote{d} \\
       \hline 
       $(\sfrac{7}{2})^+_{3}$  & - & $(\sfrac{9}{2})^-_1$ & $^{4}$I, $^{2}$H & 4401 &  -- & -- \\
       $(\sfrac{9}{2})^+_1$ & $^{4}$I, $^{2}$H & $(\sfrac{9}{2})^-_1$ & $^{4}$I, $^{2}$H & 8366 & 3405 & 3405  \\
       $(\sfrac{5}{2})^+_1$ & $^{4}$G, $^{2}$F & $(\sfrac{3}{2})^-_1$ & $^{4}$F, $^{2}$D &  44410  & 4295--4309 & 4407 - 6519 \\
    \hline \hline
    \end{tabular}
    \egroup
    \begin{tablenotes}[flushleft]
    \item[a] Leading terms for $5f^2 6d^1$ levels from  \citet{Seijo.etal:03}.
    \item[b] Leading terms for $5f^3$ levels from \citet{Roy.Prasad:20}.
    \item[c] Experimental values from \citet{Carnall.Crosswhite:85,Andres.etal:96}.
    \item[d] \textit{Ab initio} calculations from \citet{ruiperez.etal:2007,Roy.Prasad:20}.
    \end{tablenotes}
\end{threeparttable}
\end{table*}

We verify this by considering the contribution of W\,\textsc{iii} to the total emission at 4.75 and 5.22 micron, shown in Table \ref{tab:WIII_emission}. This emission consists of the photons emitted before radiative transfer is conducted, and physically corresponds to the emergent spectrum in a completely optically thin medium. Therefore, transfer effects such as absorption and scattering are not taken into account in this `optically thin' spectrum, but these effects are expected to be mostly negligible at IR wavelengths where the ejecta are optically thin. It should be noted that the contribution of individual transitions to the W\,\textsc{iii} emission at a given wavelength cannot be extracted. For instance, the $n = 4 \rightarrow 3$ ground state transition in the \textsc{fac} data has a wavelength of 5.29 micron, such that it contributes to the recorded W\,\textsc{iii} flux at 5.22 micron.

From this analysis, we find that the contribution of W\,\textsc{iii} at 4.75 micron is relatively minor ($< 10$ per cent) in all models, while somewhat more significant, though still subdominant, at 5.22 micron. Placing these lines at their correct wavelengths of 4.432 and 4.545$\mu$m, we obtain scaled A-values of 0.498\,$\rm{s^{-1}}$ and 0.534\,$\rm{s^{-1}}$ respectively, only slightly lower than the values of $\sim 0.6\,\rm{s^{-1}}$ found when considering pure LS coupling \citep[][]{Hotokezaka.etal:22}. Therefore, these lines likely emit slightly more at their true wavelengths, but may be in competition with other species emitting at the same wavelength, e.g. Se\,\textsc{iii}. Since we do not have significant optical depth that may suppress photons from escaping at these wavelengths, a more prominent W\,\textsc{iii} signal around $\sim 4 - 5$ micron in the emergent spectra likely requires a higher abundance of this species. This criteria could potentially be met in models with lower energy depositions than those yielded by dominant $\alpha$-decay, and therefore yielding higher ion fractions of W\,\textsc{iii} in low density ejecta. 

In contrast, U\,\textsc{iv} dominates emission in the B16B model at 4.5 micron over a broad range of epochs. The verification of the features arising from U\,\textsc{iv} is more involved, as no level data is readily available from NIST. The three features that are found to be significant in our model spectra arise from transitions between low-lying even parity $\rm{5f^{2}6d^{1}}$ states, to odd parity $\rm{5f^{3}}$ states. Several experimental measurements \citep{Carnall.Crosswhite:85,Andres.etal:96} and \textit{Ab initio} theoretical calculations \citep[][]{ruiperez.etal:2007,Roy.Prasad:20} of low-lying $\rm{5f^{3}}$ levels in U\,\textsc{iv} are available, while only a single theoretical calculation for $\rm{5f^{2}6d^{1}}$ levels was found \citep[][]{Seijo.etal:03}. 

One caveat of matching energy levels across different works is that it is often difficult to identify exactly which level corresponds to which due to differences in term labelling. The \textit{LSJ} term notation is more commonly used in databases and in literature, particularly for low-energy structures which are easier to compute \citep{NIST_ASD}. While \textsc{fac}, like other relativistic atomic codes, proceeds with calculations under a \textit{jj}-coupling scheme, contrary to other codes \textsc{fac} uniquely lacks an internal mechanism for converting to \textit{LSJ} notation. To address this, we employ the \textsc{jj2lsj} module \citep{Gaigalas.etal:17} from the \textsc{grasp2018} software package \citep{GRASP2}. This module transforms atomic state functions from a \textit{jj}-coupled configuration state function (CSF) basis to an \textit{LSJ}-coupled CSF basis. 
 
We use this program to extract LS leading terms from our theoretical data, and find the closest matching levels from literature. Of relevance to the transitions of interest in our model, our calculations show leading terms $^{4}\text{I}$ and $^{4}\text{F}$ for levels $(\sfrac{9}{2})^-_1$ and $(\sfrac{3}{2})^-_1$ respectively, associated with the odd $\rm{5f^3}$ configuration, and $^{2}\text{G}$, $^{4}\text{G}$ and $^{4}\text{I}$ for levels $(\sfrac{7}{2})^+_3$, $(\sfrac{5}{2})^+_1$, and $(\sfrac{9}{2})^+_1$, associated with the even $\rm{5f^2\,6d^1}$ configuration. For most levels of interest, the labels we find seem to be reliable, with the contribution of the leading CSF being greater than 64 per cent, except for level $(\sfrac{7}{2})^+_3$ where we see high mixing with first and second components of the wave-function, with contributions of 20 per cent (term $^{2}\text{G}$)  and 18 per cent (term $^{2}\text{F}$) respectively.  The wavelengths of our key transitions compared to those found using the energy levels in literature are shown in Table \ref{tab:UIV_transitions}.

From this analysis, we see that the U\,\textsc{iv} feature at 4401~$\ang$ cannot be verified as no clear matching upper level is found in literature. For the other two transitions, energy levels in literature predict much bluer wavelengths than those found in the \textsc{fac} data. By following the dipole transition A-value scaling as described above, we also find that the radiative transition rates of the transitions initially at 8366 and 44410~$\ang$ will be greatly enhanced, which suggests that U\,\textsc{iv} should be playing a larger role in the absorption of blue photons in our model. Experimental measurements of the absorption spectra of U\,\textsc{iv} for $\rm{5f^{3}}$ to $\rm{5f^{2}6d^{1}}$ transitions have also found significant transitions in wavelengths ranging from 2500 -- 6700~$\ang$ \citep[][]{Karbowiak:05}. This implies that the main impact of U\,\textsc{iv} on KN spectra may actually be strong suppression of blue photons, with potentially enhanced fluorescence to redder wavelengths. 

\begin{figure*}
    \centering
    \includegraphics[trim={0.4cm 0cm 0.4cm 0.2cm},width=0.48\linewidth]{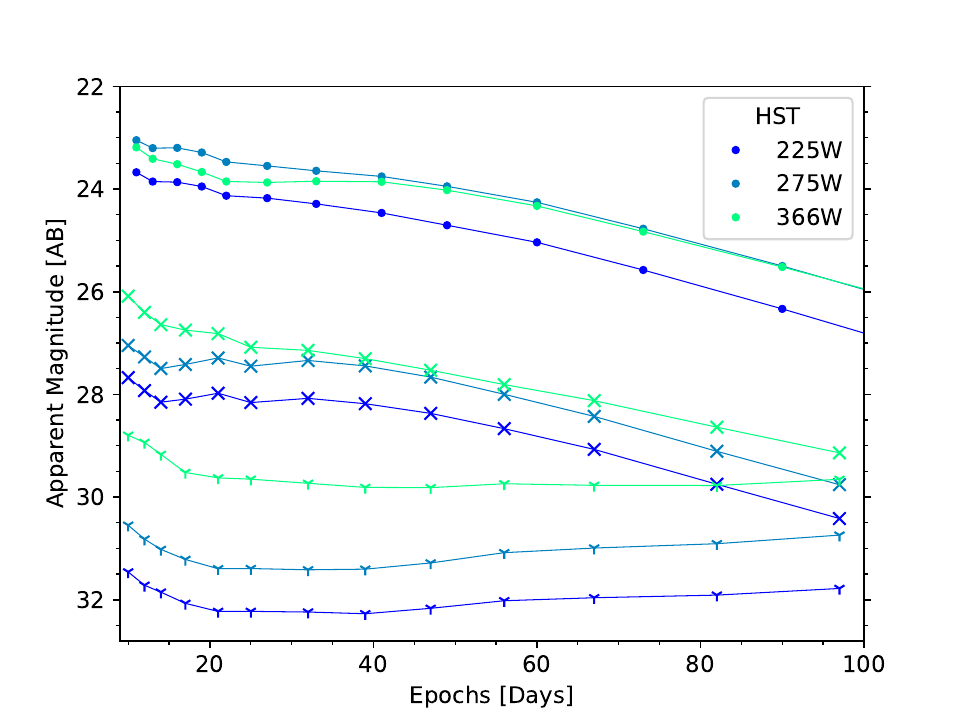}
    \includegraphics[trim={0.4cm 0cm 0.4cm 0.2cm},width=0.48\linewidth]{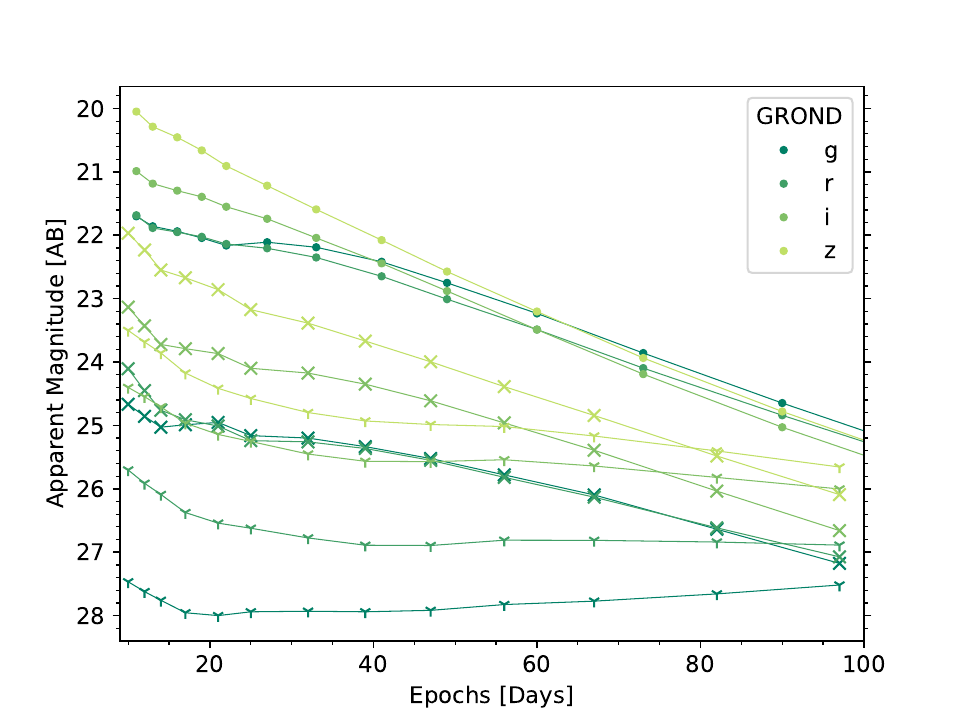}
    \includegraphics[trim={0.4cm 0cm 0.4cm 0.2cm},width=0.48\linewidth]{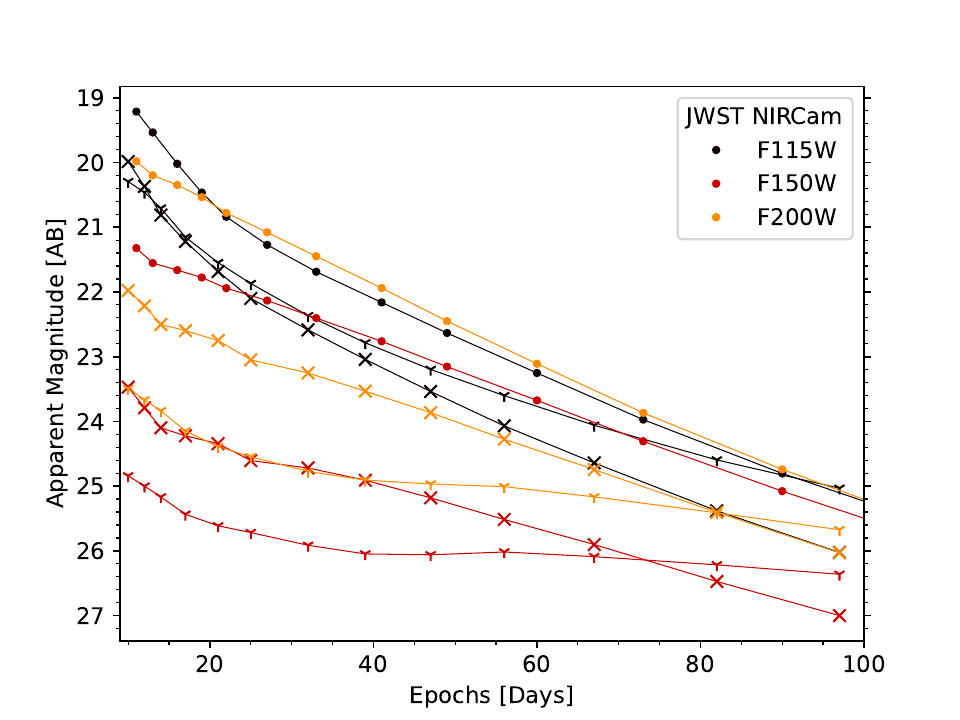}
    \includegraphics[trim={0.4cm 0cm 0.4cm 0.2cm},width=0.48\linewidth]{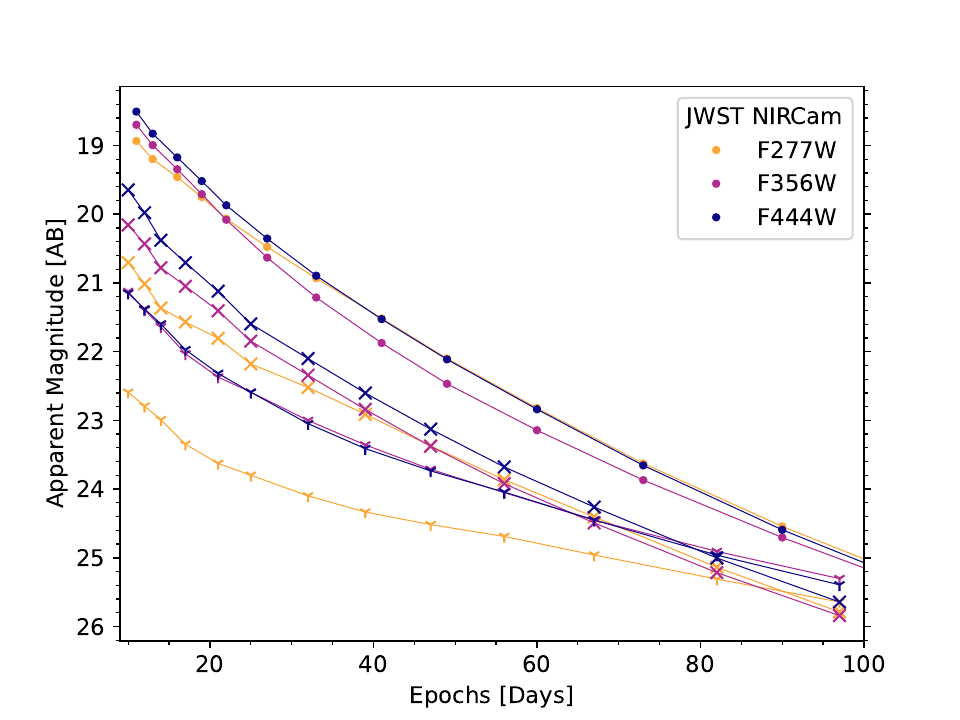}
    \caption{Broadband lightcurves of the models in the UV (top left panel), optical (top right panel) and IR (bottom panels). The points correspond to the B16A model, the crosses to the B16B model, and the `Y'-shaped markers to the W14 model.}
    \label{fig:LCs}
\end{figure*}

In general, the above analysis highlights the difficulty of robustly establishing spectral actinide signatures due to the sparsity of accurate available data. Comparison to such data when possible reveals that \textit{Ab initio} calculations from theoretical atomic physics codes that focus on completeness rather than accuracy, as is the case for the \textsc{fac} dataset used here, may lead to highly inaccurate transition wavelengths. This effect is particularly important for IR transitions, where very slight deviations in energy level structure leads to large variations in transition wavelength. However, even literature values may find significantly different energy level values depending on the calculation \citep[see e.g.][for the $\rm{5f^{3}}$ levels of U\,\textsc{iv}]{ruiperez.etal:2007,Roy.Prasad:20}. In order to provide better constraints on the spectral signatures of actinides in KN ejecta, higher accuracy atomic data is required, whether by experimental measurement or more detailed, targeted \textit{Ab initio} theoretical calculations. 


\subsection{Broadband lightcurves}
\label{subsec:LC_evolution}

In general, the B16A and B16B models predict several key actinide features that produce significant impacts on the emergent spectra. As previously noted however, the accuracy of these features is for the most part difficult to verify, with the exception of the late time Ac\,\textsc{iii} emission. Therefore, a complementary approach to the identification of actinide signatures in KNe may be taken by studying the broadband LCs of the models, which are less susceptible to the exact wavelengths of key features, and generally more representative of the overall SED shape and evolution. It should be kept in mind, however, that in the case of highly inaccurate features, such as those for Th\,\textsc{iii} in the \textsc{fac} data, the usage of photometry still suffers from this underlying inaccuracy. In this case, the magnitude of the bands around 3 micron where Th\,\textsc{iii} is most prominently emitting, particularly in the B16B model, may be somewhat overestimated. However, this is not expected to overly change the general evolution of the broadband LCs.

We conduct synthetic photometry on the emergent spectral series from near UV wavelengths to $5 \mu$m in the IR, the results of which are shown in Fig. \ref{fig:LCs}. The top-left panel shows the near UV bands, for which we have used the \textit{Hubble Space Telescope (HST)} wide-field camera's F225W, F275W and F336W filters. The top-right panel shows optical bands, where we have used the \textit{g,r,i,z} bands of the Gamma-Ray Burst Optical/Near-Infrared Detector (GROND) instrument, mounted on the MPI/ESO 2.2m telescope at La Silla in Chile\footnote{\url{https://www.mpe.mpg.de/~jcg/GROND/}}. Finally, the bottom panels shows IR bands, where the $1 - 5 \mu$m range is covered using the \textit{JWST}'s NIRCam wide bands. These choices are motivated by the spectral range of our radiative transfer simulations, which have been set to 5 micron in order to cover the 3.6 and 4.5 micron bands of \textit{Spitzer}'s IRAC, of relevance to late time observations of AT2017gfo \citep{Kasliwal.etal:22}. We have opted to use \textit{JWST}'s NIRCam for the IR photometry however, as this represents the cutting edge instrument that has and will be used for IR KNe observations in the future.

Analysing first the overall trends in each waveband, we find that the actinide-free model of W14 generally evolves qualitatively differently to the actinide-rich models. This model shows an initial decline in most wavebands, followed by a relatively flat evolution, or even a slight \textit{increase} in apparent magnitude for optical and UV wavebands. This is most noticeable for bluer wavebands, and corresponds to the spectra becoming progressively bluer with time due to rising temperatures and decreasing optical depths in homogeneous composition models \citep[][]{Pognan.etal:23}. The reddest bands ($\lambda \gtrsim 2\mu$m), however, evolve in qualitatively similar ways across all models, with the actinide-free model having a somewhat slower decline past 40 days. 

Aside from these reddest bands, the actinide-rich B16A and B16B models show a relatively different evolution. After an initial decrease, the UV and bluer optical bands tend to show a slight bump, or at least a plateau, between 20 - 40 days. Following this, a steady decrease in magnitude is found across all wavebands, the exact rate varying from band to band. This bump or plateau found in some of the broadband LCs of the actinide-rich models likely arises from flux moving bluewards as time progresses. However, in contrast to the W14 model which prolongs this plateau, or even yields a brightening in bluer wavebands, the actinide-rich models fade in every waveband past the $\sim 40$ day mark. 

In both the B16A and B16B cases, this rate of decline is greater than that of the W14 model, such that the latter appears brighter than the B16B model in certain wavebands at the end of the timespan studied here (e.g. \textit{JWST} bands from $\sim$80 days onwards). The B16A model typically remains brighter in every waveband however, likely due to the overall greater power output of the nuclear network. The generally hotter temperatures from greater radioactive power of the actinide-rich models leads them to be comparatively more blue than the W14 model. This effect is somewhat opposite of that expected by consideration of expansion opacities \citep[e.g.][]{Flors.etal:23,Fontes.etal:23}, however it is not fully clear how the red-to-blue evolution will present itself in cases of inhomogeneous composition, 3D models. As such, we present this comparison fully in the context of the models studied here.

The key differences in the evolution of the broadband LCs between the actinide-rich and free models is likely linked not only to the different model compositions, but also to the different temperature evolutions. Looking at Fig. \ref{fig:thermo}, we see that the actinide-rich models start to cool around 40 - 60 days, whereas the W14 model continues to become hotter with time. The combination of this temperature increase with the dropping optical depth at bluer wavelengths allows the actinide-free model to efficiently emit blue photons that can escape the ejecta more easily. In the actinide-rich models however, the dropping temperatures work against this process, such that it is less efficient. This factor, combined with the overall decreasing luminosity as nuclear energy input drops, leads to the steady fading of the B16A/B models across every waveband. 

These results indicate that actinide-rich and actinide-free models may be distinguished by the evolving shape of their broadband LCs, not only arising from different compositions, but perhaps mainly from different resultant nuclear power and therefore corresponding temperature evolution. From Figure \ref{fig:LCs}, we suggest that the near UV or optical LCs may provide the best possibility of establishing the significant presence or absence of actinide-bearing ejecta, given that the model LCs behave increasingly similarly the longer the wavelengths considered, such that their evolution is qualitatively similar in wavebands past $\sim 2\mu$m. The caveat here is that the near UV and optical bands are typically fainter than the IR bands, which may present a limitation on the application of this method to distant or relatively faint KNe.

\section{Discussion and Conclusion}
\label{sec:discusssion}

In this study, we have confirmed that uncertainties in nuclear physics inputs, such as mass models and nuclear decay rates, as well as choice of thermodynamic expansion history for nucleosynthesis, play important roles in the final composition and radioactive power of KNe, as has been shown in diverse prior studies \citep[e.g.][]{Barnes.etal:21,Zhu.etal:21,Kullmann.etal:23}. Using trajectories with $Y_e = 0.15$ obtained from merger simulations, and based on the radioactive power and decay products predicted by full nuclear network calculations, we have evolved three different models over a timespan of 10--100~days, producing spectra and LCs using the \textsc{sumo} NLTE radiative transfer code. 

From the emergent spectra and LCs, we find markedly different evolutions between the two actinide-rich B16A/B models, and the actinide-free W14 model. We find that the difference in radioactive power between these models leads to significantly distinct temperature and ionization structures, which drive dissimilar emergent spectral series. This effect, alongside important differences in composition, lead to noticeably different observable signatures both from a spectral and LC perspective.

The actinide-rich models evolve differently in the NLTE regime than predicted by previous results \citep[][]{Hotokezaka.etal:20,Hotokezaka.etal:21,Pognan.etal:22a}, notably undergoing a temperature downturn in the 40 - 60 day range. This originates from the dominance of $\alpha$-decay following an exponential decay law over ensemble $\beta$-decay following a $t^{-2.8}$ power law in the energy deposition. The late time cooling plays a major role in the overall evolution of these models, driving the aforementioned differences between them, and the actinide-free W14 model. Furthermore, the B16A/B models suggest actinide signatures that may be detectable in an emergent spectrum, despite the presence of other line-rich species such as lanthanides. In both cases, the actinide mass fraction remains relatively small ($X_{\rm{Ac}} \sim 0.02 - 0.03$) in comparison to the lanthanide fraction ($X_{\rm{La}} \sim$ 0.12). 

We provide a first attempt at constraining possible actinide signatures in KN spectra across a broad range of post-photospheric epochs and wavelengths. We find that our actinide atomic data as calculated by \textsc{fac} targeting completeness over accuracy, yields transitions that are for the most part highly inaccurate, or unverifiable due to lack of reference data in the literature. The exceptions to this are two transitions from Ac\,\textsc{iii}, which we find to be highly accurate, but the features of which remain subdominant in our model. Our U\,\textsc{iv} transitions are found to be too red, with experimental sources suggesting that this species should be playing a larger role in the absorption and reprocessing of blue photons by fluorescence. By considering our key Th\,\textsc{iii} transitions at their real wavelengths in the mid-IR, and scaling the A-values for these transitions following the dipole scaling rule, we find that they may be particularly good observational targets for \textit{JWST}. Further studies with improved atomic data, ejecta models, and covering the entire spectral range of \textit{JWST} are required in order to determine if low lying E1 transitions of Th\,\textsc{iii} are robust observable spectral actinide signatures.

Given the limitations of establishing the exact accuracy of actinide signatures in spectra, we also consider the photometric evolution of low $Y_e$ ejecta by examining broadband LCs. Here, we show that for a given electron fraction, actinide-bearing ejecta evolve qualitatively differently to actinide-free ejecta across most wavelengths, from the near UV to the IR up to about 2 micron. This is found to be importantly linked to the differing temperature evolution of the ejecta, as well as compositional variations. 

Generally, for a given mass, we also find that actinide-bearing ejecta may be brighter than actinide-free ejecta, due to the important presence of efficiently thermalizing $\alpha$-decay products, and in even more neutron rich ejecta ($Y_e < 0.15$), potentially also spontaneous fission fragments. These differences, detectable even without spectral data, may allow observational constraints on the presence of actinide species, and even beyond, to be established. More advanced ejecta models and parameter-space studies are required in order to provide tighter constraints, such as the mass fraction of actinides present. This initial study, however, offers an alternative way to distinguish between the presence of line rich species, such as lanthanides and actinides, that does not rely on expansion opacity calculations, which may be challenging to apply in late time, NLTE settings \citep[e.g.][]{Tanaka.etal:20,Pognan.etal:22b,Flors.etal:23}.

Combining higher accuracy atomic data for actinide structure, as well as improved ejecta models and NLTE physics treatment, subsequent studies may build on this initial work in order to provide reliable predictions of actinide signatures in NS merger ejecta. The usage of such models to confirm the presence, or absence, of actinides in future KN detections may help further establish the astrophysical origin of actinide elements, as well as provide tighter constraints on r-process sites in the Universe.

\section*{Acknowledgements}

The authors thank S. Sim for useful discussion and comments on this work, as well as the anonymous referee for thorough and helpful comments. The radiative transfer simulations were enabled by resources provided by the National Academic Infrastructure for Supercomputing in Sweden (NAISS), and the Swedish National Infrastructure for Computing (SNIC),
at the Parallelldatorcentrum (PDC) Center for High Performance
Computing, Royal Institute of Technology (KTH), partially funded by the Swedish Research Council through grant agreement no. 2022-06725. 
MRW acknowledges supports from the National Science and Techonology Council, Taiwan under Grant No.~111-2628-M-001-003-MY4, the Academia Sinica under Project No.~AS-CDA-109-M11, and the Physics Division of the
National Center for Theoretical Sciences, Taiwan. GMP and AF acknowledge support by the European Research Council (ERC) under the European Union's Horizon 2020 research and innovation programme (ERC Advanced Grant KILONOVA No.~885281), the Deutsche Forschungsgemeinschaft (DFG, German Research Foundation) - Project-ID 279384907 - SFB 1245, and MA 4248/3-1, and the State of Hesse within the Cluster Project ELEMENTS. AJ acknowledges funding by the European Research Council
(ERC) under the European Union’s Horizon 2020 Research and
Innovation Program (ERC Starting Grant 803189 – SUPERSPEC), the Swedish Research Council (grant 2018-03799), and the Knut and Alice Wallenberg Foundation ("Gravity Meets Light").
RFS acknowledges the support from National funding by FCT (Portugal), through the individual research grant 2022.10009.BD and through project funding 2022.06730.PTDC, "Atomic inputs for kilonovae modeling (ATOMIK)".
J.G. acknowledges support from the Swedish Research Council through the individual project grant with contract no. 2020-05467.

\section*{Data Availability}

The data underlying this article will be shared on reasonable request to the corresponding author.



\bibliographystyle{mnras}
\bibliography{biblio} 




\appendix

\section{Thermal collision strength scaling factors}
\label{app:thermal_coll}

The scaling factors applied to the thermal collision strength calculations based on the work of \citet{Bromley.etal:23} are shown here in Table \ref{tab:coll_strengths}.

\begin{table*}
    \centering
    \caption{Scaling factors applied to allowed transitions ($n_{\rm{regemorter}}$) and forbidden transitions ($n_{\rm{axelrod}}$) depending on ion stage and temperature solution.}
    \bgroup
    \def\arraystretch{1.2}
    \begin{tabular}{ccccccccccccc}
     \hline \hline
    \multicolumn{1}{c}{Ion Charge} & & \multicolumn{3}{c}{0} & & \multicolumn{2}{c}{1} & & \multicolumn{2}{c}{2}\\ \cline{1-1} \cline{3-5} \cline{7-8} \cline{10-11} 
       Temperature [K] & & $\leq 5000$ & $5000 < T \leq 15000$ & $> 15000$ & & $\leq 20 000$ & $> 20000 $ & & $\leq 20 000$ & $> 20000 $ \\
       \hline
        $n_{\rm{regemorter}}$ & & 10.0 & 7.5 & 5.0 & & 2.0 & 1.5 & & 1.0 & 1.0 \\
        $n_{\rm{axelrod}}$  & & 6.7 & 6.7 & 5.5 & & 5.0 & 3.5 & & 5.0 & 4.3 \\
    \hline \hline
    \end{tabular}
    \egroup
    \label{tab:coll_strengths}
\end{table*}

\section{W14 model spectra}
\label{app:W14_spectra}

\begin{figure*}
    \centering
    \includegraphics[trim={0.5cm 0cm 0.6cm 0.2cm},width=0.49\linewidth]{Fig4c.pdf}
    \includegraphics[trim={0.5cm 0cm 0.6cm 0.2cm},width=0.49\linewidth]{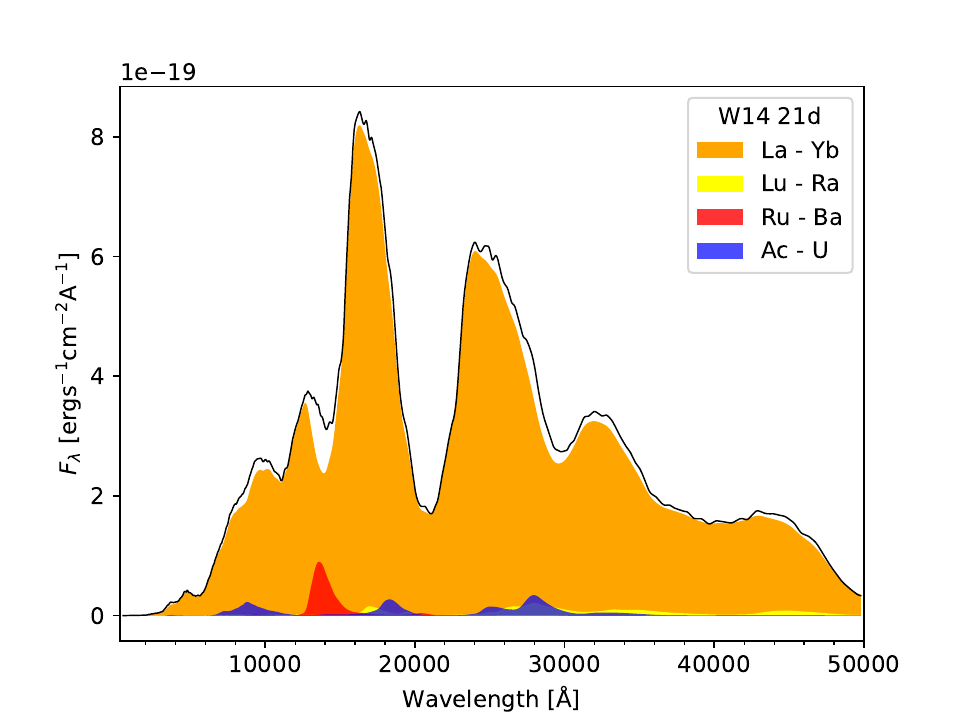}
    \includegraphics[trim={0.5cm 0cm 0.6cm 0.2cm},width=0.49\linewidth]{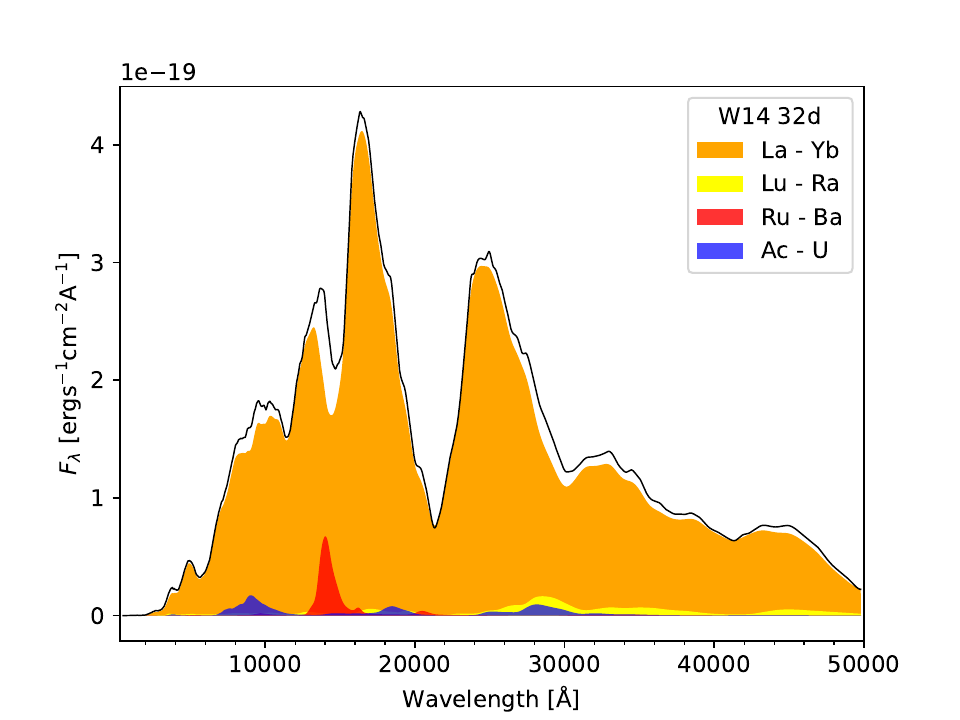}
    \includegraphics[trim={0.5cm 0cm 0.6cm 0.2cm},width=0.49\linewidth]{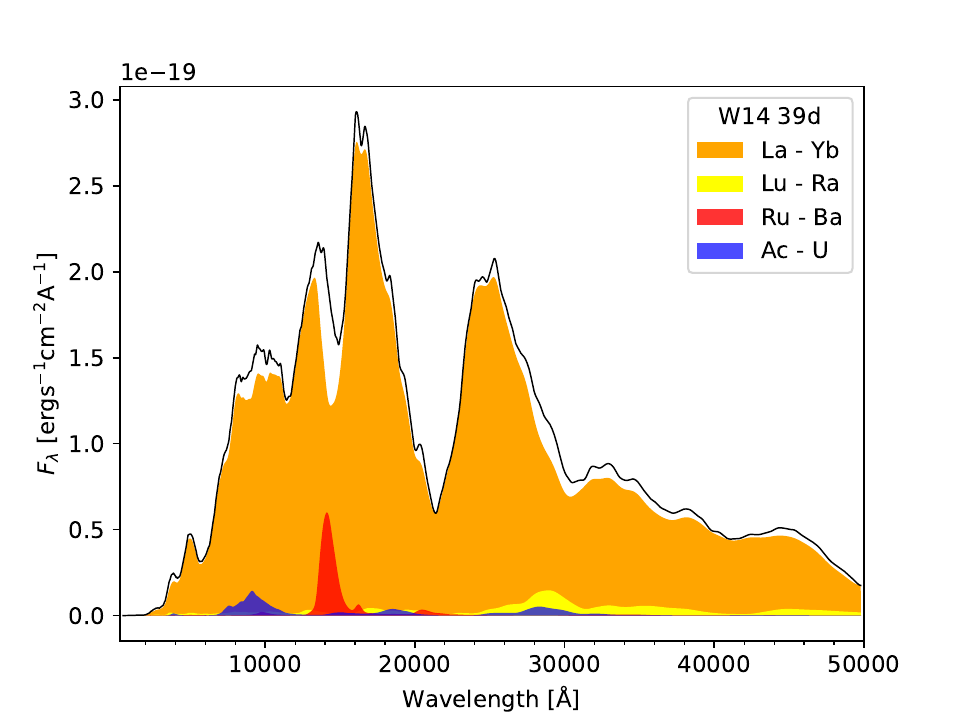}
    \includegraphics[trim={0.5cm 0cm 0.6cm 0.2cm},width=0.49\linewidth]{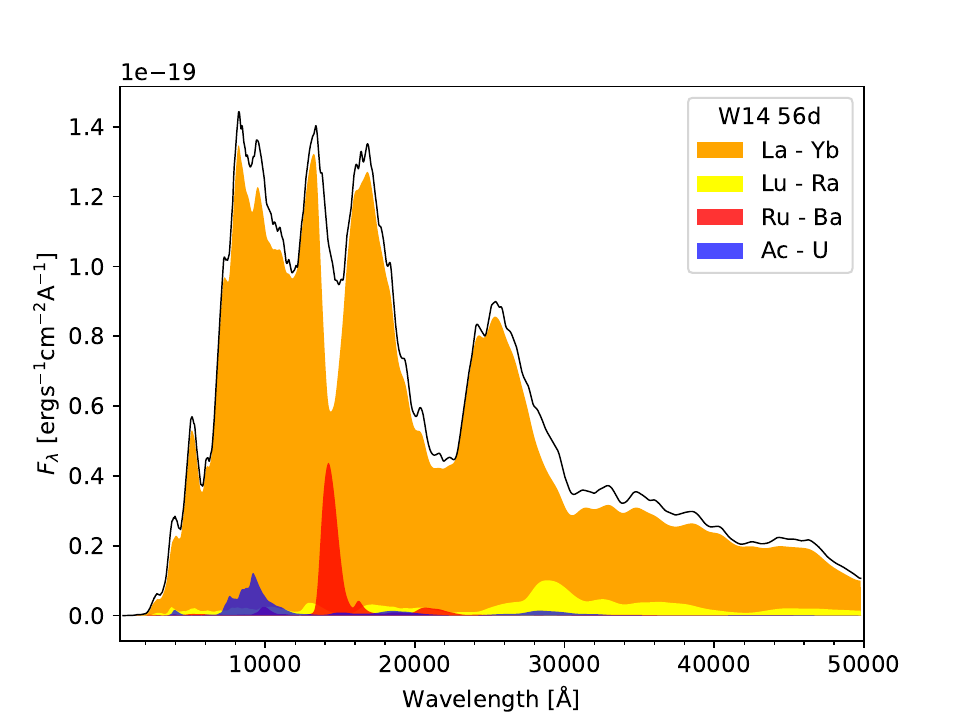}
    \includegraphics[trim={0.5cm 0cm 0.6cm 0.2cm},width=0.49\linewidth]{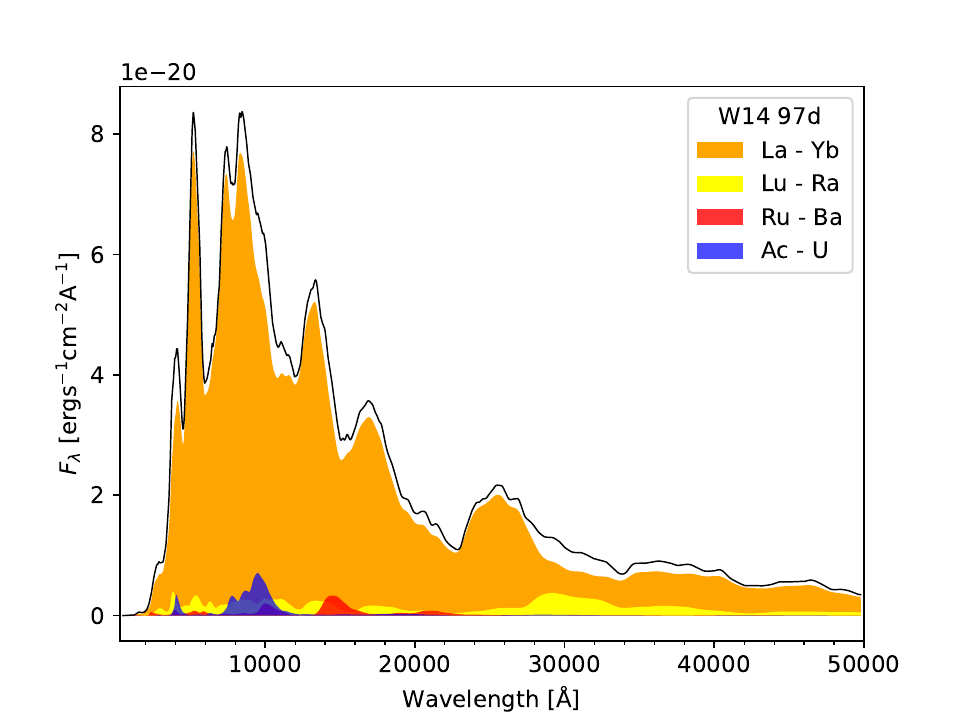}
    \caption{The W14 model spectra with element groups marked out. The contribution of actinides is found to be negligible at all epochs.}
    \label{fig:Wanajo14_spectra}
\end{figure*}

We present in Fig. \ref{fig:Wanajo14_spectra} the spectral evolution of the W14 model with the element groups marked out. As noted in the main text, the low actinide mass fraction in this model leads to negligible actinide contribution to the emergent spectra across all epochs. 

\bsp	
\label{lastpage}
\end{document}